\documentclass[pre,eqsecnum,preprint,showpacs,showkeys,preprintnumbers,amsmath,amssymb,amsfonts]{revtex4}

\input epsf

\usepackage{graphicx}
\usepackage{bm}
\usepackage{float}

\newcommand{\beq}{\begin{equation}}
\newcommand{\eeq}{\end{equation}}
\newcommand{\beqs}{\begin{eqnarray}}
\newcommand{\eeqs}{\end{eqnarray}}

\newtheorem{lemma}{Lemma}[section]
\newtheorem{defi}{Definition}[section]
\newtheorem{conj}{Conjecture}[section]
\newtheorem{propo}{Proposition}[section]

\begin{document}

\title{Dimer-monomer model on the generalized Tower of Hanoi graph}

\author{Wei-Bang Li}
\email{weibang1108@gmail.com}
\affiliation{Department of Physics \\
National Cheng Kung University \\
Tainan 70101, Taiwan}

\author{Shu-Chiuan Chang}
\email{scchang@mail.ncku.edu.tw}
\affiliation{Department of Physics \\
National Cheng Kung University \\
Tainan 70101, Taiwan}

\date{\today}

\begin{abstract}

We study the number of dimer-monomers $M_d(n)$ on the Tower of Hanoi graphs $TH_d(n)$ at stage $n$ with dimension $d$ equal to 3 and 4. The entropy per site is defined as $z_{TH_d}=\lim_{v \to \infty} \ln M_d(n)/v$, where $v$ is the number of vertices on $TH_d(n)$. We obtain the lower and upper bounds of the entropy per site, and the convergence of these bounds approaches to zero rapidly when the calculated stage increases. The numerical value of $z_{TH_d}$ is evaluated to more than a hundred digits correct. Using the results with $d$ less than or equal to 4, we predict the general form of the lower and upper bounds for $z_{TH_d}$ with arbitrary $d$.



\end{abstract}

\maketitle

\section{Introduction}
\label{sectionI}

The dimer-monomer model is an interesting but elusive model in statistical mechanics \cite{gaunt69,Heilmann70,Heilmann72}. In this model, a dimer is realized by a diatomic molecule with two neighboring sites attaching to a surface or lattice. For the sites that are not occupied by any dimers, they could be regarded as covered by monomers. Let us define $N_{DM}(G)$ to be the number of dimer-monomers on a graph $G$. 

The computation of the general dimer-monomer model remains to be a difficult problem \cite{jerrum}, in contrast to the closed-packed dimer problem on planar lattices that had been discussed and solved more than fifty years ago \cite{kasteleyn61, temperley61, fisher61}. Recent computation of close-packed dimers, dimers with a single monomer, and general dimer-monomer models on regular lattices are given in Refs. \cite{lu99, tzeng03, izmailian03, izmailian05, yan05, yan06, kong06, izmailian06, wu06, kong06n, kong06nn}. It is also interesting to discuss the dimer-monomer problem on fractals with scaling invariance but not translational invariance. The fractals with non-integer Hausdorff dimension can be constructed from certain basic shape \cite{mandelbrot,Falconer}. A famous fractal is the Tower of Hanoi graph, and it has been discussed in different contexts \cite{Angeli, distribution, Independent}. 

The dimer-monomer problem on the Tower of Hanoi graph with dimension $d=2$ was discussed in \cite{Dimer-monomer}. In this article, we shall first recall some basic definitions in section II. In section III, we present the recursion relations for the number of dimer-monomers on $TH_d(n)$ with dimension $d=3$, then enumerate the entropy per site using lower and upper bounds in details. The calculation for $TH_d(n)$ with dimension $d=4$ will be given in section IV. In the last section, we shall predict the general form of the lower and upper bounds of the entropy per site for dimer-monomers on the Tower of Hanoi graph with arbitrary dimension.

\section{Preliminaries}
\label{sectionII}

In this section, let us review some basic terminology. A graph $G=(V,E)$ that is connected and has no loops is defined by the vertex (site) set $V$ and edge (bond) set $E$ \cite{bbook,fh}. Denote $v(G)=|V|$ as the number of vertices in $G$ and $e(G)=|E|$ as the number of edges.  Two vertices $a$ and $b$ are neighboring if the edge $ab$ is included in $E$. A matching of a graph $G$ is an independent edge subset where the edges have no common vertices. The number of matching in $G$ is denoted as $N_{DM}(G)$, which corresponds to the number of dimer-monomers in statistical mechanics. Although monomer and dimer weights can be associated to each monomer and dimer (cf. \cite{wu06}), we shall set such weights to 1 here. 

$N_{DM}(G)$ can increase exponentially as the number of vertices approaches to infinity, and the entropy per site $z_G$ is defined as
\beq
z_G = \lim_{v(G) \to \infty} \frac{\ln N_{DM}(G)}{v(G)} \ ,
\label{zdef}
\eeq
where the subscript $G$ indicates the thermodynamic limit. 

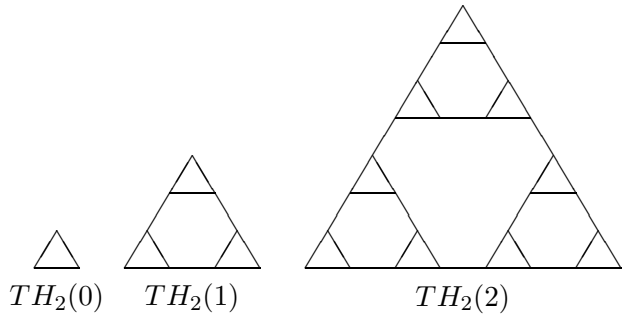
\begin{figure}[htbp]
\unitlength 1mm 
\begin{picture}(78,35)
\put(0,0){\line(1,0){6}}
\put(0,0){\line(3,5){3}}
\put(6,0){\line(-3,5){3}}
\put(3,-4){\makebox(0,0){$TH_2(0)$}}
\put(12,0){\line(1,0){18}}
\put(12,0){\line(3,5){9}}
\put(30,0){\line(-3,5){9}}
\put(18,10){\line(1,0){6}}
\put(24,0){\line(3,5){3}}
\put(18,0){\line(-3,5){3}}
\put(21,-4){\makebox(0,0){$TH_2(1)$}}
\put(36,0){\line(1,0){42}}
\put(36,0){\line(3,5){21}}
\put(78,0){\line(-3,5){21}}
\put(48,20){\line(1,0){18}}
\put(60,0){\line(3,5){9}}
\put(54,0){\line(-3,5){9}}
\multiput(42,10)(24,0){2}{\line(1,0){6}}
\multiput(48,0)(24,0){2}{\line(3,5){3}}
\multiput(42,0)(24,0){2}{\line(-3,5){3}}
\put(54,30){\line(1,0){6}}
\put(60,20){\line(3,5){3}}
\put(54,20){\line(-3,5){3}}
\put(57,-4){\makebox(0,0){$TH_2(2)$}}
\end{picture}

\vspace*{5mm}
\caption{\footnotesize{The first three stages $0 \le n \le 2$ of the two-dimensional Tower of Hanoi graph $TH_2(n)$.}} 
\label{sgfig}
\end{figure}

The two-dimensional Tower of Hanoi graph $TH_2(n)$ at stage $n=0, 1, 2$ shown in Fig. \ref{sgfig} has been discussed in Ref. \cite{Dimer-monomer}. At stage $n=0$, $TH_2(0)$ is a regular triangle. $TH_2(n+1)$ is consisted of three $TH_2(n)$ using three edges to connect the outmost vertices. Such arrangement can be generalized to construct the Tower of Hanoi graph with higher dimension. For the general Tower of Hanoi graph $TH_d(n)$, the number of edges is   
\beq
e(TH_d(n)) = {d+1 \choose 2}\frac{[(d+1)^{n+1}-1]}{d} = \frac {(d+1)}{2} [(d+1)^{n+1}-1] \ ,
\label{e}
\eeq
while the number of vertices is
\beq
v(TH_d(n)) = (d+1)^{n+1} \ .
\label{v}
\eeq
The $(d+1)$ outmost vertices of $TH_d(n)$ have degree $d$, while the other vertices have degree $(d+1)$. Therefore, in the thermodynamic limit $TH_d$ is $(d+1)$-regular.

\section{The entropy per site for dimer-monomers on $TH_3(n)$}
\label{sectionIII}

We shall consider the entropy per site for dimer-monomers on the three-dimensional Tower of Hanoi graph $TH_3(n)$ in details. The following quantities will be used in this section.

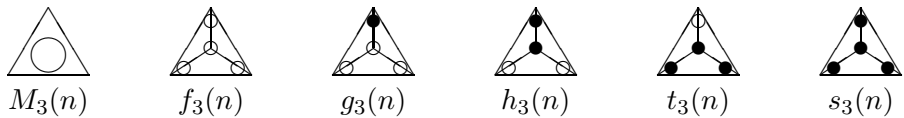
\begin{figure}[htbp]
\unitlength 1.8mm 
\begin{picture}(66,5)

\put(0,0){\line(1,0){6}}
\put(0,0){\line(3,5){3}}
\put(6,0){\line(-3,5){3}}
\put(3,1.5){\circle{2.5}}
\put(3,-2){\makebox(0,0){$M_3(n)$}}

\put(12,0){\line(1,0){6}}
\put(12,0){\line(3,5){3}}
\put(18,0){\line(-3,5){3}}
\put(12,0){\line(3,2){3}}
\put(18,0){\line(-3,2){3}}
\put(15,2){\line(0,1){3}}
\put(13,0.5){\circle{1}}
\put(17,0.5){\circle{1}}
\put(15,2){\circle{1}}
\put(15,4){\circle{1}}
\put(15,-2){\makebox(0,0){$f_3(n)$}}

\put(24,0){\line(1,0){6}}
\put(24,0){\line(3,5){3}}
\put(30,0){\line(-3,5){3}}
\put(24,0){\line(3,2){3}}
\put(30,0){\line(-3,2){3}}
\put(27,2){\line(0,1){3}}
\put(25,0.5){\circle{1}}
\put(29,0.5){\circle{1}}
\put(27,2){\circle{1}}
\put(27,4){\circle*{1}}
\put(27,-2){\makebox(0,0){$g_3(n)$}}

\put(36,0){\line(1,0){6}}
\put(36,0){\line(3,5){3}}
\put(42,0){\line(-3,5){3}}
\put(36,0){\line(3,2){3}}
\put(42,0){\line(-3,2){3}}
\put(39,2){\line(0,1){3}}
\put(37,0.5){\circle{1}}
\put(41,0.5){\circle{1}}
\put(39,2){\circle*{1}}
\put(39,4){\circle*{1}}
\put(39,-2){\makebox(0,0){$h_3(n)$}}

\put(48,0){\line(1,0){6}}
\put(48,0){\line(3,5){3}}
\put(54,0){\line(-3,5){3}}
\put(48,0){\line(3,2){3}}
\put(54,0){\line(-3,2){3}}
\put(51,2){\line(0,1){3}}
\put(49,0.5){\circle*{1}}
\put(53,0.5){\circle*{1}}
\put(51,2){\circle*{1}}
\put(51,4){\circle{1}}
\put(51,-2){\makebox(0,0){$t_3(n)$}}

\put(60,0){\line(1,0){6}}
\put(60,0){\line(3,5){3}}
\put(66,0){\line(-3,5){3}}
\put(60,0){\line(3,2){3}}
\put(66,0){\line(-3,2){3}}
\put(63,2){\line(0,1){3}}
\put(61,0.5){\circle*{1}}
\put(65,0.5){\circle*{1}}
\put(63,2){\circle*{1}}
\put(63,4){\circle*{1}}
\put(63,-2){\makebox(0,0){$s_3(n)$}}

\end{picture}

\vspace*{5mm}
\caption{\footnotesize{Illustration of $M_3(n)$, $f_3(n)$, $g_3(n)$, $h_3(n)$, $t_3(n)$, $s_3(n)$. We only show the four outmost vertices explicitly for $f_3(n)$, $g_3(n)$, $h_3(n)$, $t_3(n)$, $s_3(n)$, where each open circle is covered by a monomer while each solid circle is covered by a dimer.}} 
\label{fghtfig}
\end{figure}

\begin{defi} \label{defith3} Consider the three-dimensional Tower of Hanoi graph $TH_3(n)$ at stage $n$. (i) Define $M_{3}(n) \equiv N_{DM}(TH_{3}(n))$ to be the number of dimer-monomers. (ii) Define $f_{3}(n)$ to be the number of dimer-monomers so that all four outmost vertices are covered by monomers. (iii) Define $g_{3}(n)$ to be the number of dimer-monomers so that one certain outmost vertex (e.g. the topmost vertex shown in Fig. \ref{fghtfig}) is covered by a dimer and the other three outmost vertices are covered by monomers. (iv) Define $h_{3}(n)$ to be the number of dimer-monomers so that two certain outmost vertices (e.g. the downmost vertices shown in Fig. \ref{fghtfig}) are covered by monomers and the other two outmost vertices are covered by dimers. (v) Define $t_{3}(n)$ to be the number of dimer-monomers so that one certain outmost vertex (e.g. the topmost vertex shown in Fig. \ref{fghtfig}) is covered by a monomer and the other three outmost vertices are covered by dimers. (vi) Define $s_{3}(n)$ to be the number of dimer-monomers so that all four outmost vertices are covered by dimers.
\end{defi}

These quantities $M_3(n)$, $f_3(n)$, $g_3(n)$, $h_3(n)$, $t_3(n)$, and $s_3(n)$ are illustrated in Fig. \ref{fghtfig}, where we only show the outmost vertices explicitly. Due to rotational symmetry, there are four orientations of $g_3(n)$, six orientations of $h_3(n)$ and four orientations of $t_3(n)$, so that
\beq
M_3(n) = f_3(n)+4g_3(n)+6h_3(n)+4t_3(n)+s_3(n) 
\label{meq3}
\eeq
for a non-negative integer $n$. The values of these quantities at $n=0$ are $f_3(0)=1$, $g_3(0)=0$, $h_3(0)=1$, $t_3(0)=0$, and $s_3(n)=3$, so that $M_3(0)=10$. The aim of this section is devoted to the asymptotic behavior of $M_3(n)$. The six quantities $M_3(n)$, $f_3(n)$, $g_3(n)$, $h_3(n)$, $t_3(n)$, $s_3(n)$ satisfy the recursion relations given in the following Lemma, and they shall be written as $M$, $f$, $g$, $h$, $t$, and $s$ for simplicity in this section.

\bigskip

\begin{figure}[htbp]
\unitlength 1.8mm 
\begin{picture}(66,5)

\put(0,0){\line(1,0){6}}
\put(0,0){\line(3,5){3}}
\put(3,2){\line(0,1){3}}
\put(0,0){\line(3,2){3}}
\put(3,4){\circle{1}}
\put(6,0){\line(-3,2){3}}
\put(6,0){\line(-3,5){3}}

\put(3,-2){\makebox(0,0){$P_3(n)$}}

\put(12,0){\line(1,0){6}}
\put(12,0){\line(3,5){3}}
\put(18,0){\line(-3,5){3}}
\put(12,0){\line(3,2){3}}
\put(18,0){\line(-3,2){3}}
\put(15,2){\line(0,1){3}}
\put(13,0.5){\circle{1}}
\put(15,4){\circle{1}}
\put(15,-2){\makebox(0,0){$Q_3(n)$}}

\put(24,0){\line(1,0){6}}
\put(24,0){\line(3,5){3}}
\put(30,0){\line(-3,5){3}}
\put(24,0){\line(3,2){3}}
\put(30,0){\line(-3,2){3}}
\put(27,2){\line(0,1){3}}
\put(25,0.5){\circle{1}}
\put(29,0.5){\circle{1}}
\put(27,4){\circle{1}}
\put(27,-2){\makebox(0,0){$R_3(n)$}}

\put(36,0){\line(1,0){6}}
\put(36,0){\line(3,5){3}}
\put(42,0){\line(-3,5){3}}
\put(36,0){\line(3,2){3}}
\put(42,0){\line(-3,2){3}}
\put(39,2){\line(0,1){3}}
\put(39,4){\circle*{1}}
\put(39,-2){\makebox(0,0){$X_3(n)$}}

\put(48,0){\line(1,0){6}}
\put(48,0){\line(3,5){3}}
\put(54,0){\line(-3,5){3}}
\put(48,0){\line(3,2){3}}
\put(54,0){\line(-3,2){3}}
\put(51,2){\line(0,1){3}}
\put(49,0.5){\circle{1}}
\put(51,4){\circle*{1}}
\put(51,-2){\makebox(0,0){$Y_3(n)$}}

\put(60,0){\line(1,0){6}}
\put(60,0){\line(3,5){3}}
\put(66,0){\line(-3,5){3}}
\put(60,0){\line(3,2){3}}
\put(66,0){\line(-3,2){3}}
\put(63,2){\line(0,1){3}}
\put(61,0.5){\circle{1}}
\put(65,0.5){\circle{1}}
\put(63,4){\circle*{1}}
\put(63,-2){\makebox(0,0){$W_3(n)$}}
\end{picture}

\caption{\footnotesize{Illustration of $P_3(n)$, $Q_3(n)$, $R_3(n)$, $X_3(n)$, $Y_3(n)$, $W_3(n)$, where each open circle is covered by a monomer and each solid circle is covered by a dimer. No open circle or solid circle means that the outmost vertex can be covered by either a dimer or a monomer.}} 
\label{pqrxyw}
\end{figure}
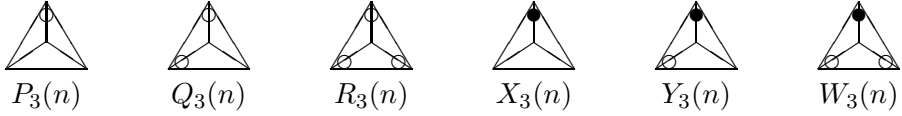

We shall define six additional quantities $P_3(n)$, $Q_3(n)$, $R_3(n)$, $X_3(n)$, $Y_3(n)$, and $W_3(n)$ as follows. Let $P_3(n)$ be the number of dimer-monomers on $TH_3(n)$ so that one certain outmost vertex is covered by a monomer, and the other three outmost vertices can be covered by either dimers or monomers. The other quantities $Q_3(n)$, $R_3(n)$, $X_3(n)$, $Y_3(n)$, and $W_3(n)$ can be defined similarly as shown in Fig. \ref{pqrxyw}, where no open circle or solid circle is put on the outmost vertex if it can be covered by either a dimer or a monomer. It is clear that
\beqs
P_3(n)&=&f_3(n)+3g_3(n)+3h_3(n)+t_3(n) \ , \cr
Q_3(n)&=&f_3(n)+2g_3(n)+h_3(n) \ , \cr
R_3(n)&=&f_3(n)+g_3(n) \ , \cr
X_3(n)&=&g_3(n)+3h_3(n)+3t_3(n)+s_3(n) \ , \cr
Y_3(n)&=&g_3(n)+2h_3(n)+t_3(n) \ , \cr
W_3(n)&=&g_3(n)+h_3(n) \ .
\label{eqspqrxyw}
\eeqs

\begin{lemma} \label{lemmasg2r} For any non-negative integer $n$, the recursion relations are given by
\beqs
f_3(n+1)& = &64f^4+384f^3g+192f^3h+32f^3t+960f^2g^2+312f^2h^2+12f^2t^2+1056f^2gh\cr
& &+192f^2gt+120f^2ht+1152fg^3+304fh^3+4ft^3+2064fg^2h+408fg^2t\cr
& &+1320fgh^2+204fh^2t+60fgt^2+48fht^2+552fght+552g^4+138h^4+t^4\cr
& &+1416g^3h+304g^3t+708gh^3+144h^3t+12gt^3+12ht^3+660g^2ht+516gh^2t\cr
& &+132ght^2+1452g^2h^2+78g^2t^2+60h^2t^2\ , 
\label{feq}
\eeqs
\beqs
g_3(n+1)& = &64f^3g+96f^3h+48f^3t+8f^3s+288f^2g^2+264f^2h^2+30f^2t^2+624f^2gh\cr
& &+288f^2gt+48f^2gs+204f^2ht+30f^2hs+6f^2ts+480fg^3+330fh^3+12ft^3\cr
& &+1392fg^2h+612fg^2t+102fg^2s+1188fgh^2+366fh^2t+51fh^2s+144fgt^2\cr
& &+117fht^2+3ft^2s+924fght+138fghs+30fgts+24fhts+288g^4+177h^4\cr
& &+3t^4+1068g^3h+456g^3t+76g^3s+802gh^3+267h^3t+36h^3s+34gt^3+33ht^3\cr
& &+t^3s+1092g^2ht+165g^2hs+39g^2ts+912gh^2t+129gh^2s+30h^2ts+309ght^2\cr
& &+9gt^2s+9ht^2s+66ghts+1392g^2h^2+180g^2t^2+141h^2t^2\ , 
\label{geq}
\eeqs
\beqs
h_3(n+1)& = &64f^2g^2+160f^2h^2+52f^2t^2+2f^2s^2+192f^2gh+96f^2gt+16f^2gs+176f^2ht\cr
& &+32f^2hs+20f^2ts+192fg^3+344fh^3+34ft^3+736fg^2h+368fg^2t+64fg^2s\cr
& &+928fgh^2+508fh^2t+92fh^2s+260fgt^2+244fht^2+16ft^2s+10fgs^2\cr
& &+8fhs^2+2fts^2+960fght+176fghs+100fgts+88fhts+160g^4+242h^4\cr
& &+10t^4+752g^3h+376g^3t+68g^3s+928gh^3+464h^3t+86h^3s+94gt^3+94ht^3\cr
& &+6t^3s+1304g^2ht+244g^2hs+130g^2ts+1364gh^2t+254gh^2s+116h^2ts\cr
& &+652ght^2+46gt^2s+46ht^2s+22ghs^2+6gts^2+6hts^2+240ghts+1292g^2h^2\cr
& &+336g^2t^2+13g^2s^2+323h^2t^2+10h^2s^2+t^2s^2\ ,
\label{heq}
\eeqs
\beqs
t_3(n+1)& = &64fg^3+288fh^3+76ft^3+fs^3+288fg^2h+144fg^2t+24fg^2s+480fgh^2\cr
& &+516fh^2t+102fh^2s+156fgt^2+330fht^2+51ft^2s+6fgs^2+15fhs^2+9t^2s^2\cr
& &+12fts^2+528fght+96fghs+60fgts+138fhts+96g^4+354h^4+36t^4\cr
& &+528g^3h+272g^3t+48g^3s+1068gh^3+802h^3t+165h^3s+228gt^3+267ht^3\cr
& &+30t^3s+3gs^3+3hs^3+ts^3+1236g^2ht+234g^2hs+144g^2ts+1824gh^2t\cr
& &+366gh^2s+297h^2ts+1092ght^2+153gt^2s+174ht^2s+51ghs^2+36gts^2\cr
& &+39hts^2+462ghts+1128g^2h^2+360g^2t^2+15g^2s^2+696h^2t^2+33h^2s^2\ ,
\label{teq}
\eeqs
\beqs
\label{seq}
s_3(n+1)& = &64g^4+552h^4+138t^4+s^4+384g^3h+192g^3t+32g^3s+1152gh^3+1416h^3t\cr
& &+304h^3s+304gt^3+708ht^3+144t^3s+4gs^3+12hs^3+12ts^3+1056g^2ht\cr
& &+192g^2hs+120g^2ts+2064gh^2t+408gh^2s+660h^2ts+1320ght^2+204gt^2s\cr
& &+516ht^2s+60ghs^2+48gts^2+132hts^2+552ghts+960g^2h^2+312g^2t^2\cr
& &+12g^2s^2+1452h^2t^2+78h^2s^2+60t^2s^2\ ,
\eeqs
\beqs
\label{meq}
M_3(n)&=&64f^4+144f^2ts+355t^4+4266h^4+3112g^4+4fs^3+2820fgt^2+150g^2s^2\cr
&&+270h^2s^2+224f^3t+3252g^2hs+1632g^2ts+3912gh^2s+2664h^2ts+1128gt^2s\cr
&&+1524ht^2s+2496f^2g^2+84fgs^2+108fhs^2+60fts^2+2328f^2h^2+12f^2s^2\cr
&&+396ghs^2+228gts^2+324hts^2+4704f^2gh+1920f^2gt+1992f^2ht+13200fg^2h\cr
&&+5640fg^2t+13560fgh^2+6780fh^2t+3300fht^2+304t^3s+18852g^2ht\cr
&&+21708gh^2t+10968ght^2+288f^2gs+312f^2hs+16gs^3+24hs^3+888fg^2s\cr
&&+1164fh^2s+312ft^2s+1176fhts+4104ghts+12696g^3h+12120fght\cr
&&+1992fghs+960fgts+640f^3g+576f^3h+1624h^3s+936g^3s+32f^3s\cr
&&+6798h^2t^2+4566g^2t^2+1928gt^3+2484ht^3+20244g^2h^2+s^4+8620h^3t\cr
&&+5664g^3t+14908gh^3+560ft^3+16ts^3+4840fh^3+444f^2t^2+4480fg^3\cr
&&+102t^2s^2\ . 
\eeqs
\end{lemma}

{\sl Proof} \quad 
The Tower of Hanoi graph $TH_3(n+1)$ is composed of four $TH_3(n)$ with six additional connecting edges. When a certain connecting edge is contained in the matching, i.e. occupied by a dimer, of $TH_3(n+1)$, its vertices in the original $TH_3(n)$ must be occupied by monomers. Let us categorized the number of dimer-monomers on $TH_3(n+1)$ in terms of the number of six additional edges contained in the matching.

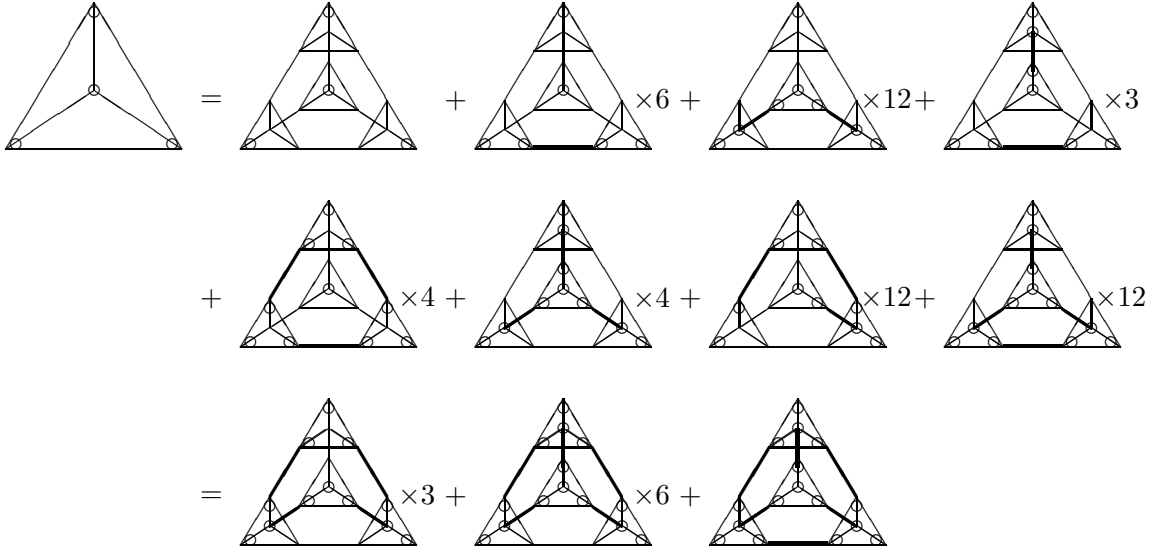
\begin{figure}[htbp]
\unitlength 1.3mm 
\begin{picture}(120,20)
\put(0,0){\line(1,0){18}}
\put(0,0){\line(3,5){9}}
\put(18,0){\line(-3,5){9}}
\put(0,0){\line(3,2){9}}
\put(18,0){\line(-3,2){9}}
\put(9,6){\line(0,1){9}}
\put(9,6){\circle{1}}
\put(1,0.5){\circle{1}}
\put(17,0.5){\circle{1}}
\put(9,14){\circle{1}}

\put(21,5){\makebox(0,0){$=$}}

\put(24,0){\line(1,0){18}}
\put(24,0){\line(3,5){9}}
\put(36,0){\line(3,5){3}}
\put(30,0){\line(-3,5){3}}
\put(42,0){\line(-3,5){9}}
\put(24,0){\line(3,2){3}}
\put(36,0){\line(3,2){3}}
\put(42,0){\line(-3,2){3}}
\put(30,0){\line(-3,2){3}}
\put(27,2){\line(0,1){3}}
\put(39,2){\line(0,1){3}}
\put(30,4){\line(1,0){6}}
\put(30,10){\line(1,0){6}}
\put(30,4){\line(3,5){3}}
\put(36,4){\line(-3,5){3}}
\put(30,4){\line(3,2){3}}
\put(30,10){\line(3,2){3}}
\put(36,4){\line(-3,2){3}}
\put(36,10){\line(-3,2){3}}
\put(33,6){\line(0,1){3}}
\put(33,12){\line(0,1){3}}
\put(27,2){\line(3,2){3}}
\put(39,2){\line(-3,2){3}}
\put(33,9){\line(0,1){3}}

\put(25,0.5){\circle{1}}

\put(33,14){\circle{1}}

\put(41,0.5){\circle{1}}

\put(33,6){\circle{1}}

\put(46,5){\makebox(0,0){$+$}}

\put(48,0){\line(1,0){18}}
\put(48,0){\line(3,5){9}}
\put(60,0){\line(3,5){3}}
\put(54,0){\line(-3,5){3}}
\put(66,0){\line(-3,5){9}}
\put(48,0){\line(3,2){3}}
\put(60,0){\line(3,2){3}}
\put(66,0){\line(-3,2){3}}
\put(54,0){\line(-3,2){3}}
\put(51,2){\line(0,1){3}}
\put(63,2){\line(0,1){3}}
\put(54,4){\line(1,0){6}}
\put(54,10){\line(1,0){6}}
\put(54,4){\line(3,5){3}}
\put(60,4){\line(-3,5){3}}
\put(54,4){\line(3,2){3}}
\put(54,10){\line(3,2){3}}
\put(60,4){\line(-3,2){3}}
\put(60,10){\line(-3,2){3}}
\put(57,6){\line(0,1){3}}
\put(57,12){\line(0,1){3}}
\put(51,2){\line(3,2){3}}
\put(63,2){\line(-3,2){3}}
\put(57,9){\line(0,1){3}}

\put(49,0.5){\circle{1}}
\put(53,0.5){\circle{1}}

\put(57,14){\circle{1}}

\put(65,0.5){\circle{1}}
\put(61,0.5){\circle{1}}

\put(57,6){\circle{1}}

\put(54,0.2){\thicklines\line(1,0){6}}

\put(66,5){\makebox(0,0){$\times 6$}}
\put(70,5){\makebox(0,0){$+$}}

\put(72,0){\line(1,0){18}}
\put(72,0){\line(3,5){9}}
\put(84,0){\line(3,5){3}}
\put(78,0){\line(-3,5){3}}
\put(90,0){\line(-3,5){9}}
\put(72,0){\line(3,2){3}}
\put(84,0){\line(3,2){3}}
\put(90,0){\line(-3,2){3}}
\put(78,0){\line(-3,2){3}}
\put(75,2){\line(0,1){3}}
\put(87,2){\line(0,1){3}}
\put(78,4){\line(1,0){6}}
\put(78,10){\line(1,0){6}}
\put(78,4){\line(3,5){3}}
\put(84,4){\line(-3,5){3}}
\put(78,4){\line(3,2){3}}
\put(78,10){\line(3,2){3}}
\put(84,4){\line(-3,2){3}}
\put(84,10){\line(-3,2){3}}
\put(81,6){\line(0,1){3}}
\put(81,12){\line(0,1){3}}
\put(75,2){\line(3,2){3}}
\put(87,2){\line(-3,2){3}}
\put(81,9){\line(0,1){3}}

\put(73,0.5){\circle{1}}
\put(75,2){\circle{1}}
\put(75,2){\thicklines\line(3,2){3}}
\put(75,1.9){\thicklines\line(3,2){3}}

\put(81,14){\circle{1}}

\put(89,0.5){\circle{1}}
\put(87,2){\circle{1}}
\put(87,2){\thicklines\line(-3,2){3}}
\put(87,1.9){\thicklines\line(-3,2){3}}

\put(81,6){\circle{1}}
\put(79,4.5){\circle{1}}
\put(83,4.5){\circle{1}}

\put(90,5){\makebox(0,0){$\times 12$}}
\put(94,5){\makebox(0,0){$+$}}

\put(96,0){\line(1,0){18}}
\put(96,0){\line(3,5){9}}
\put(108,0){\line(3,5){3}}
\put(102,0){\line(-3,5){3}}
\put(114,0){\line(-3,5){9}}
\put(96,0){\line(3,2){3}}
\put(108,0){\line(3,2){3}}
\put(114,0){\line(-3,2){3}}
\put(102,0){\line(-3,2){3}}
\put(99,2){\line(0,1){3}}
\put(111,2){\line(0,1){3}}
\put(102,4){\line(1,0){6}}
\put(102,10){\line(1,0){6}}
\put(102,4){\line(3,5){3}}
\put(108,4){\line(-3,5){3}}
\put(102,4){\line(3,2){3}}
\put(102,10){\line(3,2){3}}
\put(108,4){\line(-3,2){3}}
\put(108,10){\line(-3,2){3}}
\put(105,6){\line(0,1){3}}
\put(105,12){\line(0,1){3}}
\put(99,2){\line(3,2){3}}
\put(111,2){\line(-3,2){3}}
\put(105,9){\line(0,1){3}}

\put(97,0.5){\circle{1}}
\put(101,0.5){\circle{1}}
\put(102,0.2){\thicklines\line(1,0){6}}

\put(105,14){\circle{1}}
\put(105,12){\circle{1}}

\put(113,0.5){\circle{1}}
\put(109,0.5){\circle{1}}

\put(105,6){\circle{1}}
\put(105,8){\circle{1}}
\put(105,8){\thicklines\line(0,1){4}}
\put(105.1,8){\thicklines\line(0,1){4}}

\put(114,5){\makebox(0,0){$\times 3$}}

\end{picture}

\begin{picture}(120,20)
\put(21,5){\makebox(0,0){$+$}}
\put(24,0){\line(1,0){18}}
\put(24,0){\line(3,5){9}}
\put(36,0){\line(3,5){3}}
\put(30,0){\line(-3,5){3}}
\put(42,0){\line(-3,5){9}}
\multiput(24,0)(12,0){2}{\line(3,2){3}}
\multiput(42,0)(-12,0){2}{\line(-3,2){3}}
\multiput(27,2)(12,0){2}{\line(0,1){3}}
\multiput(30,4)(0,6){2}{\line(1,0){6}}
\put(30,4){\line(3,5){3}}
\put(36,4){\line(-3,5){3}}
\multiput(30,4)(0,6){2}{\line(3,2){3}}
\multiput(36,4)(0,6){2}{\line(-3,2){3}}
\multiput(33,6)(0,6){2}{\line(0,1){3}}
\put(27,2){\line(3,2){3}}
\put(39,2){\line(-3,2){3}}
\put(33,9){\line(0,1){3}}

\put(25,0.5){\circle{1}}
\put(27,4){\circle{1}}
\put(29,0.5){\circle{1}}
\put(27,5){\thicklines\line(3,5){3}}
\put(27,4.8){\thicklines\line(3,5){3}}
\put(30,0.2){\thicklines\line(1,0){6}}

\put(31,10.5){\circle{1}}
\put(33,14){\circle{1}}
\put(35,10.5){\circle{1}}

\put(37,0.5){\circle{1}}
\put(39,4){\circle{1}}
\put(41,0.5){\circle{1}}
\put(39,5){\thicklines\line(-3,5){3}}
\put(39,4.8){\thicklines\line(-3,5){3}}

\put(33,6){\circle{1}}

\put(46,5){\makebox(0,0){$+$}}
\put(42,5){\makebox(0,0){$\times 4$}}

\put(48,0){\line(1,0){18}}
\put(48,0){\line(3,5){9}}
\put(60,0){\line(3,5){3}}
\put(54,0){\line(-3,5){3}}
\put(66,0){\line(-3,5){9}}
\multiput(48,0)(12,0){2}{\line(3,2){3}}
\multiput(66,0)(-12,0){2}{\line(-3,2){3}}
\multiput(51,2)(12,0){2}{\line(0,1){3}}
\multiput(54,4)(0,6){2}{\line(1,0){6}}
\put(54,4){\line(3,5){3}}
\put(60,4){\line(-3,5){3}}
\multiput(54,4)(0,6){2}{\line(3,2){3}}
\multiput(60,4)(0,6){2}{\line(-3,2){3}}
\multiput(57,6)(0,6){2}{\line(0,1){3}}
\put(51,2){\line(3,2){3}}
\put(63,2){\line(-3,2){3}}
\put(57,9){\line(0,1){3}}

\put(49,0.5){\circle{1}}
\put(51,2){\circle{1}}

\put(57,14){\circle{1}}
\put(57,12){\circle{1}}

\put(65,0.5){\circle{1}}
\put(63,2){\circle{1}}

\put(55,4.5){\circle{1}}
\put(57,8){\circle{1}}
\put(59,4.5){\circle{1}}
\put(57,6){\circle{1}}

\put(51,2){\thicklines\line(3,2){3}}
\put(51,1.9){\thicklines\line(3,2){3}}
\put(63,2){\thicklines\line(-3,2){3}}
\put(63,1.9){\thicklines\line(-3,2){3}}
\put(57,8){\thicklines\line(0,1){4}}
\put(56.9,8){\thicklines\line(0,1){4}}

\put(70,5){\makebox(0,0){$+$}}
\put(66,5){\makebox(0,0){$\times 4$}}

\put(72,0){\line(1,0){18}}
\put(72,0){\line(3,5){9}}
\put(84,0){\line(3,5){3}}
\put(78,0){\line(-3,5){3}}
\put(90,0){\line(-3,5){9}}
\multiput(72,0)(12,0){2}{\line(3,2){3}}
\multiput(90,0)(-12,0){2}{\line(-3,2){3}}
\multiput(75,2)(12,0){2}{\line(0,1){3}}
\multiput(78,4)(0,6){2}{\line(1,0){6}}
\put(78,4){\line(3,5){3}}
\put(84,4){\line(-3,5){3}}
\multiput(78,4)(0,6){2}{\line(3,2){3}}
\multiput(84,4)(0,6){2}{\line(-3,2){3}}
\multiput(81,6)(0,6){2}{\line(0,1){3}}
\put(75,2){\line(3,2){3}}
\put(87,2){\line(-3,2){3}}
\put(81,9){\line(0,1){3}}

\put(73,0.5){\circle{1}}
\put(75,4){\circle{1}}
\put(75,5){\thicklines\line(3,5){3}}
\put(75,4.8){\thicklines\line(3,5){3}}

\put(79,10.5){\circle{1}}
\put(81,14){\circle{1}}
\put(83,10.5){\circle{1}}

\put(87,4){\circle{1}}
\put(89,0.5){\circle{1}}
\put(87,2){\circle{1}}
\put(87,5){\thicklines\line(-3,5){3}}
\put(87,4.8){\thicklines\line(-3,5){3}}
\put(87,2){\thicklines\line(-3,2){3}}
\put(87,1.9){\thicklines\line(-3,2){3}}

\put(83,4.5){\circle{1}}
\put(81,6){\circle{1}}

\put(94,5){\makebox(0,0){$+$}}
\put(90,5){\makebox(0,0){$\times 12$}}

\put(96,0){\line(1,0){18}}
\put(96,0){\line(3,5){9}}
\put(108,0){\line(3,5){3}}
\put(102,0){\line(-3,5){3}}
\put(114,0){\line(-3,5){9}}
\multiput(96,0)(12,0){2}{\line(3,2){3}}
\multiput(114,0)(-12,0){2}{\line(-3,2){3}}
\multiput(99,2)(12,0){2}{\line(0,1){3}}
\multiput(102,4)(0,6){2}{\line(1,0){6}}
\put(102,4){\line(3,5){3}}
\put(108,4){\line(-3,5){3}}
\multiput(102,4)(0,6){2}{\line(3,2){3}}
\multiput(108,4)(0,6){2}{\line(-3,2){3}}
\multiput(105,6)(0,6){2}{\line(0,1){3}}
\put(99,2){\line(3,2){3}}
\put(111,2){\line(-3,2){3}}
\put(105,9){\line(0,1){3}}

\put(97,0.5){\circle{1}}
\put(101,0.5){\circle{1}}
\put(99,2){\circle{1}}
\put(102,0.2){\thicklines\line(1,0){6}}
\put(99,2){\thicklines\line(3,2){3}}
\put(99,1.9){\thicklines\line(3,2){3}}

\put(105,14){\circle{1}}
\put(105,12){\circle{1}}

\put(109,0.5){\circle{1}}
\put(113,0.5){\circle{1}}
\put(111,2){\circle{1}}
\put(111,2){\thicklines\line(-3,2){3}}
\put(111,1.9){\thicklines\line(-3,2){3}}

\put(103,4.5){\circle{1}}
\put(105,8){\circle{1}}
\put(107,4.5){\circle{1}}
\put(105,6){\circle{1}}
\put(105,8){\thicklines\line(0,1){4}}
\put(104.9,8){\thicklines\line(0,1){4}}

\put(114,5){\makebox(0,0){$\times 12$}}
\end{picture}

\begin{picture}(120,20)
\put(21,5){\makebox(0,0){$=$}}
\put(24,0){\line(1,0){18}}
\put(24,0){\line(3,5){9}}
\put(36,0){\line(3,5){3}}
\put(30,0){\line(-3,5){3}}
\put(42,0){\line(-3,5){9}}
\multiput(24,0)(12,0){2}{\line(3,2){3}}
\multiput(42,0)(-12,0){2}{\line(-3,2){3}}
\multiput(27,2)(12,0){2}{\line(0,1){3}}
\multiput(30,4)(0,6){2}{\line(1,0){6}}
\put(30,4){\line(3,5){3}}
\put(36,4){\line(-3,5){3}}
\multiput(30,4)(0,6){2}{\line(3,2){3}}
\multiput(36,4)(0,6){2}{\line(-3,2){3}}
\multiput(33,6)(0,6){2}{\line(0,1){3}}
\put(27,2){\line(3,2){3}}
\put(39,2){\line(-3,2){3}}
\put(33,9){\line(0,1){3}}

\put(25,0.5){\circle{1}}
\put(27,4){\circle{1}}
\put(27,2){\circle{1}}
\put(27,2){\thicklines\line(3,2){3}}
\put(27,1.9){\thicklines\line(3,2){3}}
\put(27,5){\thicklines\line(3,5){3}}
\put(27,4.8){\thicklines\line(3,5){3}}

\put(31,10.5){\circle{1}}
\put(33,14){\circle{1}}
\put(35,10.5){\circle{1}}

\put(39,4){\circle{1}}
\put(41,0.5){\circle{1}}
\put(39,2){\circle{1}}
\put(39,2){\thicklines\line(-3,2){3}}
\put(39,5){\thicklines\line(-3,5){3}}
\put(39,1.9){\thicklines\line(-3,2){3}}
\put(39,4.8){\thicklines\line(-3,5){3}}

\put(31,4.5){\circle{1}}
\put(35,4.5){\circle{1}}
\put(33,6){\circle{1}}

\put(46,5){\makebox(0,0){$+$}}
\put(42,5){\makebox(0,0){$\times 3$}}

\put(48,0){\line(1,0){18}}
\put(48,0){\line(3,5){9}}
\put(60,0){\line(3,5){3}}
\put(54,0){\line(-3,5){3}}
\put(66,0){\line(-3,5){9}}
\multiput(48,0)(12,0){2}{\line(3,2){3}}
\multiput(66,0)(-12,0){2}{\line(-3,2){3}}
\multiput(51,2)(12,0){2}{\line(0,1){3}}
\multiput(54,4)(0,6){2}{\line(1,0){6}}
\put(54,4){\line(3,5){3}}
\put(60,4){\line(-3,5){3}}
\multiput(54,4)(0,6){2}{\line(3,2){3}}
\multiput(60,4)(0,6){2}{\line(-3,2){3}}
\multiput(57,6)(0,6){2}{\line(0,1){3}}
\put(51,2){\line(3,2){3}}
\put(63,2){\line(-3,2){3}}
\put(57,9){\line(0,1){3}}

\put(49,0.5){\circle{1}}
\put(51,4){\circle{1}}
\put(51,2){\circle{1}}
\put(51,2){\thicklines\line(3,2){3}}
\put(51,5){\thicklines\line(3,5){3}}
\put(51,1.9){\thicklines\line(3,2){3}}
\put(51,4.8){\thicklines\line(3,5){3}}

\put(55,10.5){\circle{1}}
\put(57,14){\circle{1}}
\put(59,10.5){\circle{1}}
\put(57,12){\circle{1}}

\put(63,4){\circle{1}}
\put(65,0.5){\circle{1}}
\put(63,2){\circle{1}}
\put(63,2){\thicklines\line(-3,2){3}}
\put(63,5){\thicklines\line(-3,5){3}}
\put(63,1.9){\thicklines\line(-3,2){3}}
\put(63,4.8){\thicklines\line(-3,5){3}}

\put(55,4.5){\circle{1}}
\put(57,8){\circle{1}}
\put(59,4.5){\circle{1}}
\put(57,6){\circle{1}}
\put(57,8){\thicklines\line(0,1){4}}
\put(56.9,8){\thicklines\line(0,1){4}}

\put(66,5){\makebox(0,0){$\times 6$}}
\put(70,5){\makebox(0,0){$+$}}

\put(72,0){\line(1,0){18}}
\put(72,0){\line(3,5){9}}
\put(84,0){\line(3,5){3}}
\put(78,0){\line(-3,5){3}}
\put(90,0){\line(-3,5){9}}
\multiput(72,0)(12,0){2}{\line(3,2){3}}
\multiput(90,0)(-12,0){2}{\line(-3,2){3}}
\multiput(75,2)(12,0){2}{\line(0,1){3}}
\multiput(78,4)(0,6){2}{\line(1,0){6}}
\put(78,4){\line(3,5){3}}
\put(84,4){\line(-3,5){3}}
\multiput(78,4)(0,6){2}{\line(3,2){3}}
\multiput(84,4)(0,6){2}{\line(-3,2){3}}
\multiput(81,6)(0,6){2}{\line(0,1){3}}
\put(75,2){\line(3,2){3}}
\put(87,2){\line(-3,2){3}}
\put(81,9){\line(0,1){3}}

\put(73,0.5){\circle{1}}
\put(75,4){\circle{1}}
\put(77,0.5){\circle{1}}
\put(75,2){\circle{1}}
\put(78,0.2){\thicklines\line(1,0){6}}
\put(75,2){\thicklines\line(3,2){3}}
\put(75,5){\thicklines\line(3,5){3}}
\put(75,1.9){\thicklines\line(3,2){3}}
\put(75,4.8){\thicklines\line(3,5){3}}

\put(79,10.5){\circle{1}}
\put(81,14){\circle{1}}
\put(83,10.5){\circle{1}}
\put(81,12){\circle{1}}

\put(85,0.5){\circle{1}}
\put(87,4){\circle{1}}
\put(89,0.5){\circle{1}}
\put(87,2){\circle{1}}
\put(87,5){\thicklines\line(-3,5){3}}
\put(87,2){\thicklines\line(-3,2){3}}
\put(87,4.8){\thicklines\line(-3,5){3}}
\put(87,1.9){\thicklines\line(-3,2){3}}

\put(79,4.5){\circle{1}}
\put(81,8){\circle{1}}
\put(83,4.5){\circle{1}}
\put(81,6){\circle{1}}
\put(81,8){\thicklines\line(0,1){4}}
\put(80.9,8){\thicklines\line(0,1){4}}

\end{picture}

\caption{\footnotesize{Illustration for the recursion relation of $f_3(n+1)$.}} 
\label{fpqr}
\end{figure}

The number $f_3(n+1)$ consists of the following cases. (i) ${6 \choose 0}=1$ configuration where none of the connecting edges are included in the matching, such that all four constituting $TH_3(n)$ are in the $P_3(n)$ status. (ii) ${6 \choose 1}=6$ configurations where one of the connecting edges is included in the matching, such that two $TH_3(n)$ are in the $P_3(n)$ status and the other two in the $Q_3(n)$ status. (iii) ${6 \choose 2}=15$ configurations where two of the connecting edges are included in the matching. Among these configurations, there are 12 of them such that one $TH_3(n)$ is in the $P_3(n)$ status, two in the $Q_3(n)$ status, and one in the $R_3(n)$ status. For the other 3 configurations, all four $TH_3(n)$ are in the $Q_3(n)$ status. (iv) ${6 \choose 3}=20$ configurations where three of the connecting edges are included in the matching. Among these configurations, there are 4 of them such that one $TH_3(n)$ is in the $P_3(n)$ status and the other three in the $R_3(n)$ status. There are 4 configurations such that one $TH_3(n)$ is in the $f_3(n)$ status and the other three in the $Q_3(n)$ status. For the other 12 configurations, two $TH_3(n)$ are in the $Q_3(n)$ status and the other two in the $R_3(n)$ status. (v) ${6 \choose 4}=15$ configurations where four of the connecting edges are included in the matching. Among these configurations, there are 12 of them such that one $TH_3(n)$ is in the $f_3(n)$ status, one in the $Q_3(n)$ status, and two in the $R_3(n)$ status. For the other 3 configurations, all four $TH_3(n)$ are in the $R_3(n)$ status. (vi) ${6 \choose 5}=6$ configurations where five of the connecting edges are included in the matching, such that two $TH_3(n)$ are in the $f_3(n)$ status and the other two in the $R_3(n)$ status. (vii) ${6 \choose 6}=1$ configuration where all the connecting edges are included in the matching, such that all four constituting $TH_3(n)$ are in the $f_3(n)$ status. These configuration are shown in Fig. \ref{fpqr}, so that $f_3(n+1)$ can be written as
\beqs
f_3(n+1)&=&P^4+6P^2Q^2+12PQ^2R+3Q^4+4PR^3+4fQ^3+12Q^2R^2\cr
&&+12fQR^2+3R^4+6f^2R^2+f^4 \ .
\label{fn}
\eeqs
Here we use the shorthand notations $M$, $P$, $Q$, $R$, $X$, $Y$, $W$ for $M_3(n)$, $P_3(n)$, $Q_3(n)$, $R_3(n)$, $X_3(n)$, $Y_3(n)$, $W_3(n)$ in this proof. 
Using the relations in Eq. (\ref{eqspqrxyw}) for the quantities $P_3(n)$, $Q_3(n)$, $R_3(n)$, Eq. (\ref{fn}) becomes Eq. (\ref{feq}).

By the same token, the recursion relations of $g_3(n+1)$, $h_3(n+1)$, $t_3(n+1)$, $s_3(n+1)$ can be expressed as follows.
\beqs
g_3(n+1)&=&P^3X +3PQ^2X +3P^2QY+3PQ^2W+3Q^2RX +3Q^3Y+6PQRY +R^3X\cr
&&+3fQ^2Y+6QR^2Y+3PR^2W+6Q^2RW+gQ^3 +3fR^2Y+6fQRW\cr
&&+3WR^3+3gQR^2+3f^2RW+3fgR^2+f^3g \ ,
\label{proofgeqs}
\eeqs
\beqs
h_3(n+1)&=& P^2X^2+ Q^2X^2+4PQXY+P^2Y^2+ Q^2Y^2+ 4PQYW+2PRY^2+2Q^2Y^2\cr
&&+2Q^2XW+4QRXY+2fQY^2+2gQ^2Y+8QRYW+2R^2XW+2PRW^2\cr
&&+2R^2Y^2+2Q^2W^2+3R^2W^2+2fQW^2+4fRYW+4gQRW+2gR^2Y\cr
&&+g^2R^2+f^2W^2+4fgRW+f^2g^2 \ ,
\label{proofheqs}
\eeqs
\beqs
t_3(n+1)&=& PX^3+3PXY^2+3QX^2Y+ 3PY^2W+3 RXY^2+3 QY^3+6QXYW\cr
&&+ PW^3+3 gQY^2+3RXW^2+6RY^2W+6QYW^2+ fY^3+ 3gQW^2\cr
&&+6gRYW+3fYW^2+3RW^3+3g^2RW+3fgW^2+ fg^3 \ ,
\label{proofteqs}
\eeqs
\beqs
s_3(n+1)&=& X^4+6 X^2 Y^2+3Y^4+12XY^2W+4 X W^3+ 4g Y^3+12 Y^2 W^2\cr
&&+3 W^4+12 gY W^2+6 g^2 W^2+g^4 \ .
\label{proofseqs}
\eeqs
Using the relations in Eq. (\ref{eqspqrxyw}) again, Eqs. (\ref{geq})-(\ref{seq}) are proved.

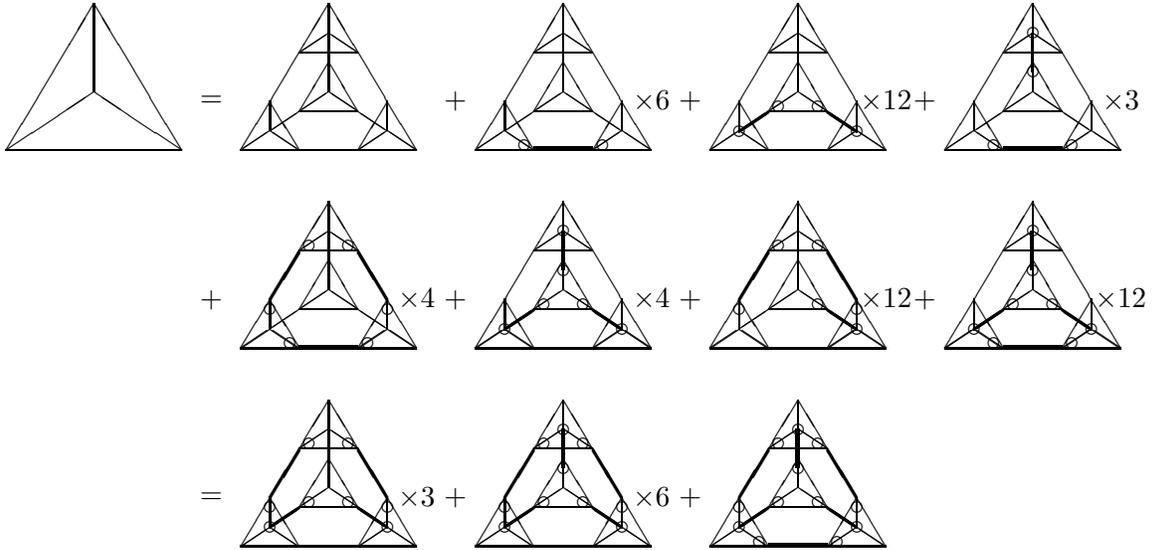
\begin{figure}[htbp]
\unitlength 1.3mm 
\begin{picture}(120,20)
\put(0,0){\line(1,0){18}}
\put(0,0){\line(3,5){9}}
\put(18,0){\line(-3,5){9}}
\put(0,0){\line(3,2){9}}
\put(18,0){\line(-3,2){9}}
\put(9,6){\line(0,1){9}}

\put(21,5){\makebox(0,0){$=$}}

\put(24,0){\line(1,0){18}}
\put(24,0){\line(3,5){9}}
\put(36,0){\line(3,5){3}}
\put(30,0){\line(-3,5){3}}
\put(42,0){\line(-3,5){9}}
\put(24,0){\line(3,2){3}}
\put(36,0){\line(3,2){3}}
\put(42,0){\line(-3,2){3}}
\put(30,0){\line(-3,2){3}}
\put(27,2){\line(0,1){3}}
\put(39,2){\line(0,1){3}}
\put(30,4){\line(1,0){6}}
\put(30,10){\line(1,0){6}}
\put(30,4){\line(3,5){3}}
\put(36,4){\line(-3,5){3}}
\put(30,4){\line(3,2){3}}
\put(30,10){\line(3,2){3}}
\put(36,4){\line(-3,2){3}}
\put(36,10){\line(-3,2){3}}
\put(33,6){\line(0,1){3}}
\put(33,12){\line(0,1){3}}
\put(27,2){\line(3,2){3}}
\put(39,2){\line(-3,2){3}}
\put(33,9){\line(0,1){3}}

\put(46,5){\makebox(0,0){$+$}}

\put(48,0){\line(1,0){18}}
\put(48,0){\line(3,5){9}}
\put(60,0){\line(3,5){3}}
\put(54,0){\line(-3,5){3}}
\put(66,0){\line(-3,5){9}}
\put(48,0){\line(3,2){3}}
\put(60,0){\line(3,2){3}}
\put(66,0){\line(-3,2){3}}
\put(54,0){\line(-3,2){3}}
\put(51,2){\line(0,1){3}}
\put(63,2){\line(0,1){3}}
\put(54,4){\line(1,0){6}}
\put(54,10){\line(1,0){6}}
\put(54,4){\line(3,5){3}}
\put(60,4){\line(-3,5){3}}
\put(54,4){\line(3,2){3}}
\put(54,10){\line(3,2){3}}
\put(60,4){\line(-3,2){3}}
\put(60,10){\line(-3,2){3}}
\put(57,6){\line(0,1){3}}
\put(57,12){\line(0,1){3}}
\put(51,2){\line(3,2){3}}
\put(63,2){\line(-3,2){3}}
\put(57,9){\line(0,1){3}}

\put(53,0.5){\circle{1}}

\put(61,0.5){\circle{1}}

\put(54,0.2){\thicklines\line(1,0){6}}

\put(66,5){\makebox(0,0){$\times 6$}}
\put(70,5){\makebox(0,0){$+$}}

\put(72,0){\line(1,0){18}}
\put(72,0){\line(3,5){9}}
\put(84,0){\line(3,5){3}}
\put(78,0){\line(-3,5){3}}
\put(90,0){\line(-3,5){9}}
\put(72,0){\line(3,2){3}}
\put(84,0){\line(3,2){3}}
\put(90,0){\line(-3,2){3}}
\put(78,0){\line(-3,2){3}}
\put(75,2){\line(0,1){3}}
\put(87,2){\line(0,1){3}}
\put(78,4){\line(1,0){6}}
\put(78,10){\line(1,0){6}}
\put(78,4){\line(3,5){3}}
\put(84,4){\line(-3,5){3}}
\put(78,4){\line(3,2){3}}
\put(78,10){\line(3,2){3}}
\put(84,4){\line(-3,2){3}}
\put(84,10){\line(-3,2){3}}
\put(81,6){\line(0,1){3}}
\put(81,12){\line(0,1){3}}
\put(75,2){\line(3,2){3}}
\put(87,2){\line(-3,2){3}}
\put(81,9){\line(0,1){3}}

\put(75,2){\circle{1}}
\put(75,2){\thicklines\line(3,2){3}}
\put(75,1.9){\thicklines\line(3,2){3}}

\put(87,2){\circle{1}}
\put(87,2){\thicklines\line(-3,2){3}}
\put(87,1.9){\thicklines\line(-3,2){3}}

\put(79,4.5){\circle{1}}
\put(83,4.5){\circle{1}}

\put(90,5){\makebox(0,0){$\times 12$}}
\put(94,5){\makebox(0,0){$+$}}

\put(96,0){\line(1,0){18}}
\put(96,0){\line(3,5){9}}
\put(108,0){\line(3,5){3}}
\put(102,0){\line(-3,5){3}}
\put(114,0){\line(-3,5){9}}
\put(96,0){\line(3,2){3}}
\put(108,0){\line(3,2){3}}
\put(114,0){\line(-3,2){3}}
\put(102,0){\line(-3,2){3}}
\put(99,2){\line(0,1){3}}
\put(111,2){\line(0,1){3}}
\put(102,4){\line(1,0){6}}
\put(102,10){\line(1,0){6}}
\put(102,4){\line(3,5){3}}
\put(108,4){\line(-3,5){3}}
\put(102,4){\line(3,2){3}}
\put(102,10){\line(3,2){3}}
\put(108,4){\line(-3,2){3}}
\put(108,10){\line(-3,2){3}}
\put(105,6){\line(0,1){3}}
\put(105,12){\line(0,1){3}}
\put(99,2){\line(3,2){3}}
\put(111,2){\line(-3,2){3}}
\put(105,9){\line(0,1){3}}

\put(101,0.5){\circle{1}}
\put(102,0.2){\thicklines\line(1,0){6}}

\put(105,12){\circle{1}}

\put(109,0.5){\circle{1}}

\put(105,8){\circle{1}}
\put(105,8){\thicklines\line(0,1){4}}

\put(114,5){\makebox(0,0){$\times 3$}}

\end{picture}

\begin{picture}(120,20)
\put(21,5){\makebox(0,0){$+$}}
\put(24,0){\line(1,0){18}}
\put(24,0){\line(3,5){9}}
\put(36,0){\line(3,5){3}}
\put(30,0){\line(-3,5){3}}
\put(42,0){\line(-3,5){9}}
\multiput(24,0)(12,0){2}{\line(3,2){3}}
\multiput(42,0)(-12,0){2}{\line(-3,2){3}}
\multiput(27,2)(12,0){2}{\line(0,1){3}}
\multiput(30,4)(0,6){2}{\line(1,0){6}}
\put(30,4){\line(3,5){3}}
\put(36,4){\line(-3,5){3}}
\multiput(30,4)(0,6){2}{\line(3,2){3}}
\multiput(36,4)(0,6){2}{\line(-3,2){3}}
\multiput(33,6)(0,6){2}{\line(0,1){3}}
\put(27,2){\line(3,2){3}}
\put(39,2){\line(-3,2){3}}
\put(33,9){\line(0,1){3}}

\put(27,4){\circle{1}}
\put(29,0.5){\circle{1}}
\put(27,5){\thicklines\line(3,5){3}}
\put(30,0.2){\thicklines\line(1,0){6}}
\put(27,4.8){\thicklines\line(3,5){3}}

\put(31,10.5){\circle{1}}
\put(35,10.5){\circle{1}}

\put(37,0.5){\circle{1}}
\put(39,4){\circle{1}}
\put(39,5){\thicklines\line(-3,5){3}}
\put(39,4.8){\thicklines\line(-3,5){3}}

\put(46,5){\makebox(0,0){$+$}}
\put(42,5){\makebox(0,0){$\times 4$}}

\put(48,0){\line(1,0){18}}
\put(48,0){\line(3,5){9}}
\put(60,0){\line(3,5){3}}
\put(54,0){\line(-3,5){3}}
\put(66,0){\line(-3,5){9}}
\multiput(48,0)(12,0){2}{\line(3,2){3}}
\multiput(66,0)(-12,0){2}{\line(-3,2){3}}
\multiput(51,2)(12,0){2}{\line(0,1){3}}
\multiput(54,4)(0,6){2}{\line(1,0){6}}
\put(54,4){\line(3,5){3}}
\put(60,4){\line(-3,5){3}}
\multiput(54,4)(0,6){2}{\line(3,2){3}}
\multiput(60,4)(0,6){2}{\line(-3,2){3}}
\multiput(57,6)(0,6){2}{\line(0,1){3}}
\put(51,2){\line(3,2){3}}
\put(63,2){\line(-3,2){3}}
\put(57,9){\line(0,1){3}}

\put(51,2){\circle{1}}

\put(57,12){\circle{1}}

\put(63,2){\circle{1}}

\put(55,4.5){\circle{1}}
\put(57,8){\circle{1}}
\put(59,4.5){\circle{1}}

\put(51,2){\thicklines\line(3,2){3}}
\put(63,2){\thicklines\line(-3,2){3}}
\put(57,8){\thicklines\line(0,1){4}}
\put(51,1.9){\thicklines\line(3,2){3}}
\put(63,1.9){\thicklines\line(-3,2){3}}
\put(56.9,8){\thicklines\line(0,1){4}}

\put(70,5){\makebox(0,0){$+$}}
\put(66,5){\makebox(0,0){$\times 4$}}

\put(72,0){\line(1,0){18}}
\put(72,0){\line(3,5){9}}
\put(84,0){\line(3,5){3}}
\put(78,0){\line(-3,5){3}}
\put(90,0){\line(-3,5){9}}
\multiput(72,0)(12,0){2}{\line(3,2){3}}
\multiput(90,0)(-12,0){2}{\line(-3,2){3}}
\multiput(75,2)(12,0){2}{\line(0,1){3}}
\multiput(78,4)(0,6){2}{\line(1,0){6}}
\put(78,4){\line(3,5){3}}
\put(84,4){\line(-3,5){3}}
\multiput(78,4)(0,6){2}{\line(3,2){3}}
\multiput(84,4)(0,6){2}{\line(-3,2){3}}
\multiput(81,6)(0,6){2}{\line(0,1){3}}
\put(75,2){\line(3,2){3}}
\put(87,2){\line(-3,2){3}}
\put(81,9){\line(0,1){3}}

\put(75,4){\circle{1}}
\put(75,5){\thicklines\line(3,5){3}}
\put(75,4.8){\thicklines\line(3,5){3}}

\put(79,10.5){\circle{1}}
\put(83,10.5){\circle{1}}

\put(87,4){\circle{1}}
\put(87,2){\circle{1}}
\put(87,5){\thicklines\line(-3,5){3}}
\put(87,2){\thicklines\line(-3,2){3}}
\put(87,4.8){\thicklines\line(-3,5){3}}
\put(87,1.9){\thicklines\line(-3,2){3}}

\put(83,4.5){\circle{1}}

\put(94,5){\makebox(0,0){$+$}}
\put(90,5){\makebox(0,0){$\times 12$}}

\put(96,0){\line(1,0){18}}
\put(96,0){\line(3,5){9}}
\put(108,0){\line(3,5){3}}
\put(102,0){\line(-3,5){3}}
\put(114,0){\line(-3,5){9}}
\multiput(96,0)(12,0){2}{\line(3,2){3}}
\multiput(114,0)(-12,0){2}{\line(-3,2){3}}
\multiput(99,2)(12,0){2}{\line(0,1){3}}
\multiput(102,4)(0,6){2}{\line(1,0){6}}
\put(102,4){\line(3,5){3}}
\put(108,4){\line(-3,5){3}}
\multiput(102,4)(0,6){2}{\line(3,2){3}}
\multiput(108,4)(0,6){2}{\line(-3,2){3}}
\multiput(105,6)(0,6){2}{\line(0,1){3}}
\put(99,2){\line(3,2){3}}
\put(111,2){\line(-3,2){3}}
\put(105,9){\line(0,1){3}}

\put(101,0.5){\circle{1}}
\put(99,2){\circle{1}}
\put(102,0.2){\thicklines\line(1,0){6}}
\put(99,2){\thicklines\line(3,2){3}}
\put(99,1.9){\thicklines\line(3,2){3}}

\put(105,12){\circle{1}}

\put(109,0.5){\circle{1}}
\put(111,2){\circle{1}}
\put(111,2){\thicklines\line(-3,2){3}}
\put(111,1.9){\thicklines\line(-3,2){3}}

\put(103,4.5){\circle{1}}
\put(105,8){\circle{1}}
\put(107,4.5){\circle{1}}
\put(105,8){\thicklines\line(0,1){4}}
\put(104.9,8){\thicklines\line(0,1){4}}

\put(114,5){\makebox(0,0){$\times 12$}}
\end{picture}

\begin{picture}(120,20)
\put(21,5){\makebox(0,0){$=$}}
\put(24,0){\line(1,0){18}}
\put(24,0){\line(3,5){9}}
\put(36,0){\line(3,5){3}}
\put(30,0){\line(-3,5){3}}
\put(42,0){\line(-3,5){9}}
\multiput(24,0)(12,0){2}{\line(3,2){3}}
\multiput(42,0)(-12,0){2}{\line(-3,2){3}}
\multiput(27,2)(12,0){2}{\line(0,1){3}}
\multiput(30,4)(0,6){2}{\line(1,0){6}}
\put(30,4){\line(3,5){3}}
\put(36,4){\line(-3,5){3}}
\multiput(30,4)(0,6){2}{\line(3,2){3}}
\multiput(36,4)(0,6){2}{\line(-3,2){3}}
\multiput(33,6)(0,6){2}{\line(0,1){3}}
\put(27,2){\line(3,2){3}}
\put(39,2){\line(-3,2){3}}
\put(33,9){\line(0,1){3}}

\put(27,4){\circle{1}}
\put(27,2){\circle{1}}
\put(27,2){\thicklines\line(3,2){3}}
\put(27,5){\thicklines\line(3,5){3}}
\put(27,1.9){\thicklines\line(3,2){3}}
\put(27,4.8){\thicklines\line(3,5){3}}

\put(31,10.5){\circle{1}}
\put(35,10.5){\circle{1}}

\put(39,4){\circle{1}}
\put(39,2){\circle{1}}
\put(39,2){\thicklines\line(-3,2){3}}
\put(39,5){\thicklines\line(-3,5){3}}
\put(39,1.9){\thicklines\line(-3,2){3}}
\put(39,4.8){\thicklines\line(-3,5){3}}

\put(31,4.5){\circle{1}}
\put(35,4.5){\circle{1}}

\put(46,5){\makebox(0,0){$+$}}
\put(42,5){\makebox(0,0){$\times 3$}}

\put(48,0){\line(1,0){18}}
\put(48,0){\line(3,5){9}}
\put(60,0){\line(3,5){3}}
\put(54,0){\line(-3,5){3}}
\put(66,0){\line(-3,5){9}}
\multiput(48,0)(12,0){2}{\line(3,2){3}}
\multiput(66,0)(-12,0){2}{\line(-3,2){3}}
\multiput(51,2)(12,0){2}{\line(0,1){3}}
\multiput(54,4)(0,6){2}{\line(1,0){6}}
\put(54,4){\line(3,5){3}}
\put(60,4){\line(-3,5){3}}
\multiput(54,4)(0,6){2}{\line(3,2){3}}
\multiput(60,4)(0,6){2}{\line(-3,2){3}}
\multiput(57,6)(0,6){2}{\line(0,1){3}}
\put(51,2){\line(3,2){3}}
\put(63,2){\line(-3,2){3}}
\put(57,9){\line(0,1){3}}

\put(51,4){\circle{1}}
\put(51,2){\circle{1}}
\put(51,2){\thicklines\line(3,2){3}}
\put(51,5){\thicklines\line(3,5){3}}
\put(51,1.9){\thicklines\line(3,2){3}}
\put(51,4.8){\thicklines\line(3,5){3}}

\put(55,10.5){\circle{1}}
\put(59,10.5){\circle{1}}
\put(57,12){\circle{1}}

\put(63,4){\circle{1}}
\put(63,2){\circle{1}}
\put(63,2){\thicklines\line(-3,2){3}}
\put(63,5){\thicklines\line(-3,5){3}}
\put(63,1.9){\thicklines\line(-3,2){3}}
\put(63,4.8){\thicklines\line(-3,5){3}}

\put(55,4.5){\circle{1}}
\put(57,8){\circle{1}}
\put(59,4.5){\circle{1}}
\put(57,8){\thicklines\line(0,1){4}}
\put(56.9,8){\thicklines\line(0,1){4}}

\put(66,5){\makebox(0,0){$\times 6$}}
\put(70,5){\makebox(0,0){$+$}}

\put(72,0){\line(1,0){18}}
\put(72,0){\line(3,5){9}}
\put(84,0){\line(3,5){3}}
\put(78,0){\line(-3,5){3}}
\put(90,0){\line(-3,5){9}}
\multiput(72,0)(12,0){2}{\line(3,2){3}}
\multiput(90,0)(-12,0){2}{\line(-3,2){3}}
\multiput(75,2)(12,0){2}{\line(0,1){3}}
\multiput(78,4)(0,6){2}{\line(1,0){6}}
\put(78,4){\line(3,5){3}}
\put(84,4){\line(-3,5){3}}
\multiput(78,4)(0,6){2}{\line(3,2){3}}
\multiput(84,4)(0,6){2}{\line(-3,2){3}}
\multiput(81,6)(0,6){2}{\line(0,1){3}}
\put(75,2){\line(3,2){3}}
\put(87,2){\line(-3,2){3}}
\put(81,9){\line(0,1){3}}

\put(75,4){\circle{1}}
\put(77,0.5){\circle{1}}
\put(75,2){\circle{1}}
\put(78,0.2){\thicklines\line(1,0){6}}
\put(75,2){\thicklines\line(3,2){3}}
\put(75,5){\thicklines\line(3,5){3}}
\put(75,1.9){\thicklines\line(3,2){3}}
\put(75,4.8){\thicklines\line(3,5){3}}

\put(79,10.5){\circle{1}}
\put(83,10.5){\circle{1}}
\put(81,12){\circle{1}}

\put(85,0.5){\circle{1}}
\put(87,4){\circle{1}}
\put(87,2){\circle{1}}
\put(87,5){\thicklines\line(-3,5){3}}
\put(87,2){\thicklines\line(-3,2){3}}
\put(87,4.8){\thicklines\line(-3,5){3}}
\put(87,1.9){\thicklines\line(-3,2){3}}

\put(79,4.5){\circle{1}}
\put(81,8){\circle{1}}
\put(83,4.5){\circle{1}}
\put(81,8){\thicklines\line(0,1){4}}
\put(80.9,8){\thicklines\line(0,1){4}}

\end{picture}

\caption{\footnotesize{Illustration for the recursion relation of $M_3(n+1)$. }} 
\label{mpqr}
\end{figure}

Finally, the number of dimer-monomer, $M_3(n+1)$, is given by  
\beqs
M_3(n+1)&=&M^4+6M^2P^2+12MP^2Q+3P^4+4MQ^3+4P^3R+12P^2Q^2\cr
&&+12PQ^2R+3Q^4+6Q^2R^2+R^4\ ,
\label{proofmeqs}
\eeqs
as illustrated in Fig. \ref{mpqr}. Using the relations in Eq. (\ref{eqspqrxyw}), Eq. (\ref{proofmeqs}) becomes Eq. (\ref{meq}), and the proof is completed.  \ $\Box$

$f_3(n)$, $g_3(n)$, $h_3(n)$, $t_3(n)$, $s_3(n)$, and $M_3(n)$ can be evaluated using Eqs. (\ref{feq})-(\ref{meq}), and their values for $n=0, 1, 2$ are listed in Table \ref{tablesg}. However, these numbers increase exponentially as $n$ increase, and have no simple integer factorizations. 

In the rest part of this section, we shall estimate the entropy per site $z_{TH_3}=\lim_{n \to \infty} \ln M_3(n)/v(TH_3(n))$. For the three-dimensional Tower of Hanoi graph, let us define the ratios
\beq
\alpha_3(n) = \frac{f_3(n)}{g_3(n)} \ , \qquad 
\beta_3(n) = \frac{g_3(n)}{h_3(n)} \ , \qquad 
\gamma_3(n) = \frac{h_3(n)}{t_3(n)} \ , \qquad 
\omega_3(n) = \frac{t_3(n)}{s_3(n)} \ ,
\label{ratiodef}
\eeq
and their values for $1 \le n \le 4$ are listed in Table \ref{tablesratio3}. From the first few values of $f_3(n)$, $g_3(n)$, $h_3(n)$, $t_3(n)$, $s_3(n)$, one can see that $f_3(n) < g_3(n) < h_3(n) < t_3(n) < s_3(n)$ when $n =1, 2$ and it is easy to prove this inequality by induction for all positive integer $n$. Therefore, we have $\alpha_3(n)$, $\beta_3(n)$, $\gamma_3(n)$, $\omega_3(n)$ $\in (0,1)$. The relationship of these ratios is given in the following Lemma.

\begin{table}[H]
\caption{\label{tablesg} The values of $f_3(n)$, $g_3(n)$, $h_3(n)$, $t_3(n)$, $s_3(n)$, $M_3(n)$ with $0 \le n \le 2$.}
\begin{center}
\begin{tabular}{|c||r|r|r|}
\hline\hline 
$n$      & 0 &        1 &                                2 \\ \hline\hline 
$f_3(n)$ & 1 &    1,010 &           49,464,202,269,253,193 \\ \hline 
$g_3(n)$ & 0 &    1,242 &           62,379,666,478,434,024 \\ \hline 
$h_3(n)$ & 1 &    1,556 &           78,668,504,245,191,833 \\ \hline 
$t_3(n)$ & 0 &    1,983 &           99,212,077,110,534,768 \\ \hline
$s_3(n)$ & 3 &    2,571 &          125,122,091,640,871,731 \\ \hline 
$M_3(n)$ & 10&   25,817 &        1,292,964,293,737,151,090 \\  
\hline\hline 
\end{tabular}
\end{center}
\end{table}

\begin{table}[H]
\caption{\label{tablesratio3} The values of $\alpha_3(n)$, $\beta_3(n)$, $\gamma_3(n)$, $\omega_3(n)$ with $1 \le n \le 4$. The last digits are rounded off.}
\begin{center}
\begin{tabular}{|c||r|r|r|r|r|}
\hline\hline 
$n$           &                 1 &                 2 &                 3 &                    4 \\ \hline\hline 
$\alpha_3(n)$ & 0.813204508856683 & 0.792953939347432 & 0.792939105706120 & 0.792939105697681 \\ \hline 
$\beta_3(n)$  & 0.798200514138817 & 0.792943339611629 & 0.792939105700090 & 0.792939105697681 \\ \hline 
$\gamma_3(n)$ & 0.784669692385275 & 0.792932741016451 & 0.792939105694060 & 0.792939105697681 \\ \hline
$\omega_3(n)$ & 0.771295215869312 & 0.792922143559552 & 0.792939105688030 & 0.792939105697681 \\ \hline\hline 
\end{tabular}
\end{center}
\end{table}

\begin{lemma} \label{lemmasg2c} For any integer $n>0$, 
the values of the ratios $\alpha_3(n)$, $\beta_3(n)$, $\gamma_3(n)$, and $\omega_3(n)$ are ordered as
\beq
0 \le \omega_3(n) \le \gamma_3(n) \le \beta_3(n) \le \alpha_3(n) \le 1 \ ,
\label{ratiosg2}
\eeq
and they are equal in the large $n$ limit
\beq
\lim_{n \rightarrow \infty}\alpha_3(n) = \lim_{n \rightarrow \infty}\beta_3(n) = \lim_{n \rightarrow \infty}\gamma_3(n) = \lim_{n \rightarrow \infty}\omega_3(n) \ .
\label{ordersg2}
\eeq
\end{lemma}

{\sl Proof} \quad 
It is obvious that all these ratios are positive because $f_3(n)$, $g_3(n)$, $h_3(n)$, $t_3(n)$, $s_3(n)$ are positive. 
Let us write $\alpha_3(n+1)$, $\beta_3(n+1)$, $\gamma_3(n+1)$ and $\omega_3(n+1)$ as
\beqs
\alpha_3(n+1)=\omega_3(n)\frac{A_3(n)}{B_3(n)} \ , \quad \beta_3(n+1)=\omega_3(n)\frac{B_3(n)}{C_3(n)} \ , \cr\cr
\gamma_3(n+1)=\omega_3(n)\frac{C_3(n)}{D_3(n)} \ , \quad \omega_3(n+1)=\omega_3(n)\frac{D_3(n)}{E_3(n)} \ , 
\label{rationp1}
\eeqs
where 
\beqs
A_3(n)&=&64\gamma^4\beta^4\alpha^4+384\gamma^4\beta^4\alpha^3+960\gamma^4\beta^4\alpha^2+1152\gamma^4\beta^4\alpha+552\gamma^4\beta^4+192\gamma^4\beta^3\alpha^3\cr
&&+1056\gamma^4\beta^3\alpha^2+2064\gamma^4\beta^3\alpha+1416\gamma^4\beta^3+312\gamma^4\beta^2\alpha^2+1320\gamma^4\beta^2\alpha+1452\gamma^4\beta^2\cr
&&+304\gamma^4\beta\alpha+708\gamma^4\beta+138\gamma^4+32\gamma^3\beta^3\alpha^3+192\gamma^3\beta^3\alpha^2+408\gamma^3\beta^3\alpha+304\gamma^3\beta^3\cr
&&+120\gamma^3\beta^2\alpha^2+552\gamma^3\beta^2\alpha+660\gamma^3\beta^2+204\gamma^3\beta\alpha+516\gamma^3\beta+144\gamma^3+12\gamma^2\beta^2\alpha^2\cr
&&+60\gamma^2\beta^2\alpha+78\gamma^2\beta^2+48\gamma^2\beta\alpha+132\gamma^2\beta+60\gamma^2+4\gamma\beta\alpha+12\gamma\beta+12\gamma+1 \ ,\cr
B_3(n)&=&64\omega\gamma^4\beta^4\alpha^3+288\omega\gamma^4\beta^4\alpha^2+480\omega\gamma^4\beta^4\alpha+288\omega\gamma^4\beta^4+96\omega\gamma^4\beta^3\alpha^3+624\omega\gamma^4\beta^3\alpha^2\cr
&&+1392\omega\gamma^4\beta^3\alpha+1068\omega\gamma^4\beta^3+264\omega\gamma^4\beta^2\alpha^2+1188\omega\gamma^4\beta^2\alpha+1392\omega\gamma^4\beta^2\cr
&&+330\omega\gamma^4\beta\alpha+802\omega\gamma^4\beta+177\omega\gamma^4+48\omega\gamma^3\beta^3\alpha^3+288\omega\gamma^3\beta^3\alpha^2+612\omega\gamma^3\beta^3\alpha\cr
&&+456\omega\gamma^3\beta^3+204\omega\gamma^3\beta^2\alpha^2+924\omega\gamma^3\beta^2\alpha+1092\omega\gamma^3\beta^2+366\omega\gamma^3\beta\alpha+912\omega\gamma^3\beta\cr
&&+267\omega\gamma^3+30\omega\gamma^2\beta^2\alpha^2+144\omega\gamma^2\beta^2\alpha+180\omega\gamma^2\beta^2+117\omega\gamma^2\beta\alpha+309\omega\gamma^2\beta\cr
&&+141\omega\gamma^2+12\omega\gamma\beta\alpha+34\omega\gamma\beta+33\omega\gamma+3\omega+8\gamma^3\beta^3\alpha^3+48\gamma^3\beta^3\alpha^2+102\gamma^3\beta^3\alpha\cr
&&+76\gamma^3\beta^3+30\gamma^3\beta^2\alpha^2+138\gamma^3\beta^2\alpha+165\gamma^3\beta^2+51\gamma^3\beta\alpha+129\gamma^3\beta+36\gamma^3\cr
&&+6\gamma^2\beta^2\alpha^2+30\gamma^2\beta^2\alpha+39\gamma^2\beta^2+24\gamma^2\beta\alpha+66\gamma^2\beta+30\gamma^2+3\gamma\beta\alpha+9\gamma\beta+9\gamma\cr
&&+1 \ ,\cr
C_3(n)&=&64\omega^2\gamma^4\beta^4\alpha^2+192\omega^2\gamma^4\beta^4\alpha+160\omega^2\gamma^4\beta^4+192\omega^2\gamma^4\beta^3\alpha^2+736\omega^2\gamma^4\beta^3\alpha\cr
&&+752\omega^2\gamma^4\beta^3+160\omega^2\gamma^4\beta^2\alpha^2+928\omega^2\gamma^4\beta^2\alpha+1292\omega^2\gamma^4\beta^2+344\omega^2\gamma^4\beta\alpha\cr
&&+928\omega^2\gamma^4\beta+242\omega^2\gamma^4+96\omega^2\gamma^3\beta^3\alpha^2+368\omega^2\gamma^3\beta^3\alpha+376\omega^2\gamma^3\beta^3+176\omega^2\gamma^3\beta^2\alpha^2\cr
&&+960\omega^2\gamma^3\beta^2\alpha+1304\omega^2\gamma^3\beta^2+508\omega^2\gamma^3\beta\alpha+1364\omega^2\gamma^3\beta+464\omega^2\gamma^3+52\omega^2\gamma^2\beta^2\alpha^2\cr
&&+260\omega^2\gamma^2\beta^2\alpha+336\omega^2\gamma^2\beta^2+244\omega^2\gamma^2\beta\alpha+652\omega^2\gamma^2\beta+323\omega^2\gamma^2+34\omega^2\gamma\beta\alpha\cr
&&+94\omega^2\gamma\beta+94\omega^2\gamma+10\omega^2+16\omega\gamma^3\beta^3\alpha^2+64\omega\gamma^3\beta^3\alpha+68\omega\gamma^3\beta^3+32\omega\gamma^3\beta^2\alpha^2\cr
&&+176\omega\gamma^3\beta^2\alpha+244\omega\gamma^3\beta^2+92\omega\gamma^3\beta\alpha+254\omega\gamma^3\beta+86\omega\gamma^3+20\omega\gamma^2\beta^2\alpha^2\cr
&&+100\omega\gamma^2\beta^2\alpha+130\omega\gamma^2\beta^2+88\omega\gamma^2\beta\alpha+240\omega\gamma^2\beta+116\omega\gamma^2+16\omega\gamma\beta\alpha+46\omega\gamma\beta\cr
&&+46\omega\gamma+6\omega+2\gamma^2\beta^2\alpha^2+10\gamma^2\beta^2\alpha+13\gamma^2\beta^2+8\gamma^2\beta\alpha+22\gamma^2\beta+10\gamma^2+2\gamma\beta\alpha\cr
&&+6\gamma\beta+6\gamma+1 \ ,\cr
D_3(n)&=&64\omega^3\gamma^4\beta^4\alpha+96\omega^3\gamma^4\beta^4+288\omega^3\gamma^4\beta^3\alpha+528\omega^3\gamma^4\beta^3+480\omega^3\gamma^4\beta^2\alpha+1128\omega^3\gamma^4\beta^2\cr
&&+288\omega^3\gamma^4\beta\alpha+1068\omega^3\gamma^4\beta+354\omega^3\gamma^4+144\omega^3\gamma^3\beta^3\alpha+272\omega^3\gamma^3\beta^3+528\omega^3\gamma^3\beta^2\alpha\cr
&&+1236\omega^3\gamma^3\beta^2+516\omega^3\gamma^3\beta\alpha+1824\omega^3\gamma^3\beta+802\omega^3\gamma^3+156\omega^3\gamma^2\beta^2\alpha+360\omega^3\gamma^2\beta^2\cr
&&+330\omega^3\gamma^2\beta\alpha+1092\omega^3\gamma^2\beta+696\omega^3\gamma^2+76\omega^3\gamma\beta\alpha+228\omega^3\gamma\beta+267\omega^3\gamma+36\omega^3\cr
&&+24\omega^2\gamma^3\beta^3\alpha+48\omega^2\gamma^3\beta^3+96\omega^2\gamma^3\beta^2\alpha+234\omega^2\gamma^3\beta^2+102\omega^2\gamma^3\beta\alpha+366\omega^2\gamma^3\beta\cr
&&+165\omega^2\gamma^3+60\omega^2\gamma^2\beta^2\alpha+144\omega^2\gamma^2\beta^2+138\omega^2\gamma^2\beta\alpha+462\omega^2\gamma^2\beta+297\omega^2\gamma^2\cr
&&+51\omega^2\gamma\beta\alpha+153\omega^2\gamma\beta+174\omega^2\gamma+30\omega^2+6\omega\gamma^2\beta^2\alpha+15\omega\gamma^2\beta^2+15\omega\gamma^2\beta\alpha\cr
&&+51\omega\gamma^2\beta+33\omega\gamma^2+12\omega\gamma\beta\alpha+36\omega\gamma\beta+39\omega\gamma+9\omega+\gamma\beta\alpha+3\gamma\beta+3\gamma+1 \ ,\cr
E_3(n)&=&64\omega^4\gamma^4\beta^4+384\omega^4\gamma^4\beta^3+960\omega^4\gamma^4\beta^2+1152\omega^4\gamma^4\beta+552\omega^4\gamma^4+192\omega^4\gamma^3\beta^3\cr
&&+1056\omega^4\gamma^3\beta^2+2064\omega^4\gamma^3\beta+1416\omega^4\gamma^3+312\omega^4\gamma^2\beta^2+1320\omega^4\gamma^2\beta+1452\omega^4\gamma^2\cr
&&+304\omega^4\gamma\beta+708\omega^4\gamma+138\omega^4+32\omega^3\gamma^3\beta^3+192\omega^3\gamma^3\beta^2+408\omega^3\gamma^3\beta+304\omega^3\gamma^3\cr
&&+120\omega^3\gamma^2\beta^2+552\omega^3\gamma^2\beta+660\omega^3\gamma^2+204\omega^3\gamma\beta+516\omega^3\gamma+144\omega^3+12\omega^2\gamma^2\beta^2\cr
&&+60\omega^2\gamma^2\beta+78\omega^2\gamma^2+48\omega^2\gamma\beta+132\omega^2\gamma+60\omega^2+4\omega\gamma\beta+12\omega\gamma+12\omega+1 \ .\cr
&&\quad
\eeqs
Here we use the shorthand notations $\alpha$, $\beta$, $\gamma$, $\omega$ for $\alpha_3(n)$, $\beta_3(n)$, $\gamma_3(n)$, $\omega_3(n)$ throughout this section. 

We shall demonstrate that $2/3 < \omega_3(n) <\gamma_3(n) < \beta_3(n) < \alpha_3(n) <1$ by induction on $n$, and prove that the differences between these ratios approach to zero as $n$ increases. The inequality $2/3 < \omega_3(n) < \gamma_3(n) < \beta_3(n) < \alpha_3(n) < 1$ holds for $1 \le n \le 3$ as shown in Table \ref{tablesratio3}. Let us assume that it remains valid for a certain positive integer $n$.

It is not hard to see that $\alpha_3(n)$ decreases while $\omega_3(n)$ increases as $n$ increases, namely, $ \alpha_3(n+1) < \alpha_3(n)$ and $ \omega_3(n) < \omega_3(n+1) $, that is shown in the appendix. Define $\epsilon_3(n) = \alpha_3(n) - \omega_3(n) $, and $0 \le \epsilon_3(n) \le 1/10$ holds for $1 \le n \le 4$. Assume that $ \epsilon_3(n) > \alpha_3(n) - \beta_3(n) $, $ \epsilon_3(n) > \beta_3(n) - \gamma_3(n) $ and $ \epsilon_3(n) > \gamma_3(n) - \omega_3(n) $ for a positive integer $n$, then
\beqs
& & \alpha_3(n+1)-\beta_3(n+1)=\omega_3(n)\frac{A_3(n)}{B_3(n)} - \omega_3(n)\frac{B_3(n)}{C_3(n)}=\omega_3(n)\frac{[ A_3(n)C_3(n) - B_3(n)^2 ]}{B_3(n)C_3(n)} \ . \cr &&
\eeqs
Define the positive quantities
\beq
a_3(n) = \alpha_3(n) - \beta_3(n) \ , \quad
b_3(n) = \beta_3(n) - \gamma_3(n) \ , \quad
c_3(n) = \gamma_3(n) - \omega_3(n) \ ,
\eeq
and write $A_3(n)C_3(n) - B_3(n)^2$ in terms of $a_3(n)$, $b_3(n)$, $c_3(n)$, $\omega_3(n)$ in descending power of $\omega_3(n)$, 
\beqs
A_3(n)C_3(n) - B_3(n)^2&=& 1024\omega_3(n)^{20} a^2+\omega_3(n)^{19}(4096a^3+8192a^2+10240a^2b+18432a^2c \cr
&&+2048ab)+\omega_3(n)^{18}(6144a^4+36864a^3b+...)+...+4bc^2+c^2 \cr
&<&\epsilon_3(n)^2\sum_{l=0}^{20}\sum_{i,j,k} \omega_3(n)^la_3(n)^ib_3(n)^jc_3(n)^k \ , \quad 0 \le l+i+j+k \le 20 \ ,\cr
&&\quad 
\eeqs
where the summation of the powers of $a_3(n)$, $b_3(n)$, $c_3(n)$ for each term is at least two, and we use the fact that $\epsilon_3(n)$ is larger than $a_3(n)$, $b_3(n)$, or $c_3(n)$. It follows that
\beqs
& & 0 < \alpha_3(n+1)-\beta_3(n+1) < \omega_3(n)\epsilon_3(n)^2\frac{\sum_{l=0}^{20}\sum_{i,j,k} \omega_3(n)^la_3(n)^ib_3(n)^jc_3(n)^k}{B_3(n)C_3(n)} < \epsilon_3(n)^2 \ , \cr & &
\eeqs
because $B_3(n)C_3(n)> \omega_3(n)\sum_{l=0}^{20}\sum_{i,j,k} \omega_3(n)^la_3(n)^ib_3(n)^jc_3(n)^k$.

By the same method, we also have $0 < \beta_3(n+1) - \gamma_3(n+1) < \epsilon_3(n)^2$, $0 < \gamma_3(n+1) - \omega_3(n+1) < \epsilon_3(n)^2$, such that
\beqs
0 & < & \epsilon_3(n+1) = \alpha_3(n+1)-\omega_3(n+1)\cr
&=&[\alpha_3(n+1) - \beta_3(n+1)]+[\beta_3(n+1) - \gamma_3(n+1)]+[\gamma_3(n+1) - \omega_3(n+1)] < 3\epsilon_3(n)^2 \ . \cr &&
\eeqs
Therefore, we have
\beqs
\epsilon_3(n) < 3\epsilon_3(n-1)^2 <3[3\epsilon_3(n-2)^2]^2 <...<\frac{1}{3}[3\epsilon_3(k)]^{2^{n-k}} 
\eeqs
for a positive integer $k \le n$.
Because $0 \le \epsilon_3(k) \le 1/10 $ is valid for a small positive integer $k$, $\epsilon_3(n)$ decreases as $n$ increases and the values $\alpha_3(n)$, $\beta_3(n)$, $\gamma_3(n)$, $\omega_3(n)$ are closed to each other when $n$ becomes large. The value of $\epsilon_3(n+1)$ is actually smaller than $\epsilon_3(n)^2$, and their ratios for $1 \le n \le 4$ are listed in Table \ref{epsilon2}.
Numerically, we find that in the large $n$ limit
\beq
\lim_{n\rightarrow\infty}\alpha_3(n)=\lim_{n\rightarrow\infty}\beta_3(n)=\lim_{n\rightarrow\infty}\gamma_3(n)=\lim_{n\rightarrow\infty}\omega_3(n)=0.79293910569768130956986961523.. 
\eeq
and the proof is completed.  \ $\Box$

\begin{table}[H]
\caption{\label{epsilon2} The values of $\epsilon_3(n+1) / \epsilon_3(n)^2$ with $1 \le n \le 4$. The last digits are rounded off.}
\begin{center}
\begin{tabular}{|c||r|r|r|r|}
\hline\hline 
$n$ &  1 &  2 &  3 &  4 \\ \hline\hline 
$\epsilon_3(n+1) / \epsilon_3(n)^2$ & 0.18102932094933 & 0.17893865402990 & 0.17893332747848 & 0.17893332747295 \\ \hline\hline
\end{tabular}
\end{center}
\end{table}

The lower and upper bounds for dimer-monomers on the three-dimensional Tower of Hanoi graph $TH_3(n)$, and the bounds for the entropy per site $z_{TH_3}$ are given in the following lemmas.

\begin{lemma} \label{lemma3}
For any positive integer $k \le n$, the number of dimer-monomer $M_3(n)$ is bounded:
\beqs
s_3(k)^{4^{n-k}}[1+2\omega_3(k)+2\omega_3(k)^2]^{2(4^{n-k}-1)}[1+\omega_3(n)]^4 < M_3(n) <\cr
s_3(k)^{4^{n-k}}[1+2\alpha_3(k)+2\alpha_3(k)^2]^{2(4^{n-k}-1)}[1+\alpha_3(n)]^4 \ .
\eeqs
\end{lemma}

{\sl Proof} \quad By Eq. (\ref{seq}) and the simple denotation, we get the upper bound of $s_3(n+1)$ as follows.
\beqs
s_3(n+1)&=&s^4(64\omega^4\gamma^4\beta^4+384\omega^4\gamma^4\beta^3+960\omega^4\gamma^4\beta^2+1152\omega^4\gamma^4\beta+552\omega^4\gamma^4+192\omega^4\gamma^3\beta^3\cr
&&+1056\omega^4\gamma^3\beta^2+2064\omega^4\gamma^3\beta+1416\omega^4\gamma^3+312\omega^4\gamma^2\beta^2+1320\omega^4\gamma^2\beta+1452\omega^4\gamma^2\cr
&&+304\omega^4\gamma\beta+708\omega^4\gamma+138\omega^4+32\omega^3\gamma^3\beta^3+192\omega^3\gamma^3\beta^2+408\omega^3\gamma^3\beta+304\omega^3\gamma^3\cr
&&+120\omega^3\gamma^2\beta^2+552\omega^3\gamma^2\beta+660\omega^3\gamma^2+204\omega^3\gamma\beta+516\omega^3\gamma+144\omega^3+12\omega^2\gamma^2\beta^2\cr
&&+60\omega^2\gamma^2\beta+78\omega^2\gamma^2+48\omega^2\gamma\beta+132\omega^2\gamma+60\omega^2+4\omega\gamma\beta+12\omega\gamma+12\omega+1)\cr
&<&s^4(1+12\alpha+72\alpha^2+280\alpha^3+780\alpha^4+1632\alpha^5+2624\alpha^6+3264\alpha^7+3120\alpha^8\cr
&&+2240\alpha^9+1152\alpha^{10}+384\alpha^{11}+64\alpha^{12})\cr
&=&s_3(n)^4(1+2\alpha_3(n)+2\alpha_3(n)^2)^6\cr
&<&[s_3(n-1)^4(1+2\alpha_3(n-1)+2\alpha_3(n-1)^2)^6]^4(1+2\alpha_3(n)+2\alpha_3(n)^2)^6\cr
&<&s_3(n-1)^{16}(1+2\alpha_3(n-1)+2\alpha_3(n-1)^2)^{6+24}<...\cr
&<&
s_3(k+1)^{4^{n-k}}(1+2\alpha_3(k+1)+2\alpha_3(k+1)^2)^{2(4^{n-k}-1)} \ ,
\eeqs
where we use the fact that $\alpha_3(n)$ is a monotonically decreasing function shown in the appendix, such that
\beqs
M_3(n)&=&f_3(n)+4g_3(n)+6h_3(n)+4t_3(n)+s_3(n)\cr
&=&s_3(n)[1+4\omega_3(n)+6\gamma_3(n)\omega_3(n)+4\beta_3(n)\gamma_3(n)\omega_3(n)+\alpha_3(n)\beta_3(n)\gamma_3(n)\omega_3(n)]\cr
&<&s_3(n)[1+4\alpha_3(n)+6\alpha_3(n)^2+4\alpha_3(n)^3+\alpha_3(n)^4]=s_3(n)[1+\alpha_3(n)]^4\cr
&<&s_3(k)^{4^{n-k}}[1+2\alpha_3(k)+2\alpha_3(k)^2]^{2(4^{n-k}-1)}[1+\alpha_3(n)]^4 \ .
\eeqs
The lower bound of $M_3(n)$ can be obtained similarly. \quad $\Box$

\begin{lemma} \label{lemma4}
The entropy per site for dimer-monomers on the Tower of Hanoi graph, $z_{TH_3}$, is bounded:
\beqs
\frac{\ln{s_3(k)}}{4^{k+1}}+\frac{2\ln{[1+2\omega_3(k)+2\omega_3(k)^2]}}{4^{k+1}}<z_{TH_3}<\frac{\ln{s_3(k)}}{4^{k+1}}+\frac{2\ln{[1+2\alpha_3(k)+2\alpha_3(k)^2]}}{4^{k+1}} \ ,\cr
\eeqs \label{zth3}
where $k$ is a positive integer and $z_{TH_3}=\lim_{n\rightarrow\infty}\frac{\ln{M_3(n)}}{v(TH_3(n))}$.
\end{lemma}

{\sl Proof} \quad Using Lemma \ref{lemma3} and $v(TH_3(n))=4^{n+1}$ in Eq. (\ref{v}), the lower and upper bounds for $z_{TH_3}$ can be easily derived as follows.
\beqs
z_{TH_3}&>&\lim_{n\rightarrow\infty}\Big\{\frac{\ln{s_3(k)}}{4^{k+1}}+\frac{2\ln{[1+2\omega_3(k)+2\omega_3(k)^2]}}{4^{k+1}}-\frac{2\ln{[1+2\omega_3(k)+2\omega_3(k)^2]}}{4^{n+1}}\cr
&&\qquad +\frac{\ln{[1+\omega_3(n)]}}{4^{n}}\Big\}\cr\cr
&=&\frac{\ln{s_3(k)}}{4^{k+1}}+\frac{2\ln{[1+2\omega_3(k)+2\omega_3(k)^2]}}{4^{k+1}} \ ,
\eeqs
where the last two terms on the right-hand-side of the inequality approach to zero in the infinite $n$ limit. Similarly,
\beqs
z_{TH_3}&<&\lim_{n\rightarrow\infty}\Big\{\frac{\ln{s_3(k)}}{4^{k+1}}+\frac{2\ln{[1+2\alpha_3(k)+2\alpha_3(k)^2]}}{4^{k+1}}-\frac{2\ln{[1+2\alpha_3(k)+2\alpha_3(k)^2]}}{4^{n+1}}\cr
&&\qquad +\frac{\ln{[1+\alpha_3(n)]}}{4^{n}}\Big\}\cr\cr
&=&\frac{\ln{s_3(k)}}{4^{k+1}}+\frac{2\ln{[1+2\alpha_3(k)+2\alpha_3(k)^2]}}{4^{k+1}} \ ,
\eeqs
and the proof is completed. \quad $\Box$

\begin{propo}
The entropy per site for dimer-monomers on the Tower of Hanoi graph with $d=3$ in the thermodynamic limit is
\beq
z_{TH_3}=0.65719921144295911522...\
\eeq
\end{propo}

The convergence of the lower and upper bounds is rapid. The value of $z_{TH_3}$ is accurate to one hundred and one decimals when $k$ in Lemma \ref{zth3} is equal to six.

\section{The entropy per site for dimer-monomers on $TH_4(n)$}
\label{sectionIV}

We shall derive the entropy per site for dimer-monomers on the four-dimensional Tower of Hanoi graph $TH_4(n)$ in this section. The  method is the same as that in the previous section.

\begin{defi} \label{defith4} Consider the four-dimensional Tower of Hanoi graph $TH_4(n)$ at stage $n$. (i) Define $M_{4}(n) \equiv N_{DM}(TH_{4}(n))$ to be the number of dimer-monomers. (ii) Define $f_{4}(n)$ to be the number of dimer-monomers so that all five outmost vertices are covered by monomers. (iii) Define $g_{4}(n)$ to be the number of dimer-monomers so that one certain outmost vertex is covered by a dimer while the other four outmost vertices are covered by monomers. (iv) Define $h_{4}(n)$ to be the number of dimer-monomers so that two certain outmost vertices are covered by dimers while the other three outmost vertices are covered by monomers. (v) Define $t_{4}(n)$ to be the number of dimer-monomers so that three certain outmost vertices are covered by dimers while the other two outmost vertices are covered by monomers. (vi) Define $s_{4}(n)$ to be the number of dimer-monomers so that one certain outmost vertex is covered by a monomer while the other four outmost vertices are covered by dimers. (vii) Define $u_{4}(n)$ to be the number of dimer-monomers so that all five outmost vertices are covered by dimers.
\end{defi}

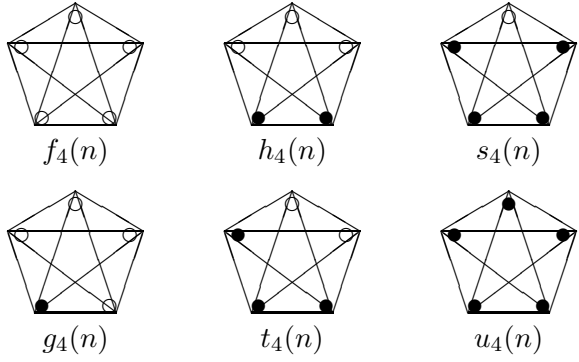
\begin{figure}[htbp]
\unitlength 1.8mm 
\begin{picture}(40,9)
\put(2,0){\line(1,0){6}}
\put(2,0){\line(4,3){8}}
\put(2,0){\line(-1,3){2}}
\put(2,0){\line(1,3){3}}
\put(8,0){\line(-1,3){3}}
\put(8,0){\line(1,3){2}}
\put(8,0){\line(-4,3){8}}
\put(0,6){\line(1,0){10}}
\put(5,9){\line(5,-3){5}}
\put(5,9){\line(-5,-3){5}}
\put(2.5,0.5){\circle{1}}
\put(7.5,0.5){\circle{1}}
\put(1,5.7){\circle{1}}
\put(9,5.7){\circle{1}}
\put(5,8){\circle{1}}
\put(5,-2){\makebox(0,0){$f_4(n)$}}
\put(18,0){\line(1,0){6}}
\put(18,0){\line(4,3){8}}
\put(18,0){\line(-1,3){2}}
\put(18,0){\line(1,3){3}}
\put(24,0){\line(-1,3){3}}
\put(24,0){\line(1,3){2}}
\put(24,0){\line(-4,3){8}}
\put(16,6){\line(1,0){10}}
\put(21,9){\line(5,-3){5}}
\put(21,9){\line(-5,-3){5}}
\put(18.5,0.5){\circle*{1}}
\put(23.5,0.5){\circle*{1}}
\put(17,5.7){\circle{1}}
\put(25,5.7){\circle{1}}
\put(21,8){\circle{1}}
\put(21,-2){\makebox(0,0){$h_4(n)$}}
\put(34,0){\line(1,0){6}}
\put(34,0){\line(4,3){8}}
\put(34,0){\line(-1,3){2}}
\put(34,0){\line(1,3){3}}
\put(40,0){\line(-1,3){3}}
\put(40,0){\line(1,3){2}}
\put(40,0){\line(-4,3){8}}
\put(32,6){\line(1,0){10}}
\put(37,9){\line(5,-3){5}}
\put(37,9){\line(-5,-3){5}}
\put(34.5,0.5){\circle*{1}}
\put(39.5,0.5){\circle*{1}}
\put(33,5.7){\circle*{1}}
\put(41,5.7){\circle*{1}}
\put(37,8){\circle{1}}
\put(37,-2){\makebox(0,0){$s_4(n)$}}
\end{picture}

\bigskip
\bigskip

\unitlength 1.8mm 
\begin{picture}(40,9)
\put(2,0){\line(1,0){6}}
\put(2,0){\line(4,3){8}}
\put(2,0){\line(-1,3){2}}
\put(2,0){\line(1,3){3}}
\put(8,0){\line(-1,3){3}}
\put(8,0){\line(1,3){2}}
\put(8,0){\line(-4,3){8}}
\put(0,6){\line(1,0){10}}
\put(5,9){\line(5,-3){5}}
\put(5,9){\line(-5,-3){5}}
\put(2.5,0.5){\circle*{1}}
\put(7.5,0.5){\circle{1}}
\put(1,5.7){\circle{1}}
\put(9,5.7){\circle{1}}
\put(5,8){\circle{1}}
\put(5,-2){\makebox(0,0){$g_4(n)$}}
\put(18,0){\line(1,0){6}}
\put(18,0){\line(4,3){8}}
\put(18,0){\line(-1,3){2}}
\put(18,0){\line(1,3){3}}
\put(24,0){\line(-1,3){3}}
\put(24,0){\line(1,3){2}}
\put(24,0){\line(-4,3){8}}
\put(16,6){\line(1,0){10}}
\put(21,9){\line(5,-3){5}}
\put(21,9){\line(-5,-3){5}}
\put(18.5,0.5){\circle*{1}}
\put(23.5,0.5){\circle*{1}}
\put(17,5.7){\circle*{1}}
\put(25,5.7){\circle{1}}
\put(21,8){\circle{1}}
\put(21,-2){\makebox(0,0){$t_4(n)$}}
\put(34,0){\line(1,0){6}}
\put(34,0){\line(4,3){8}}
\put(34,0){\line(-1,3){2}}
\put(34,0){\line(1,3){3}}
\put(40,0){\line(-1,3){3}}
\put(40,0){\line(1,3){2}}
\put(40,0){\line(-4,3){8}}
\put(32,6){\line(1,0){10}}
\put(37,9){\line(5,-3){5}}
\put(37,9){\line(-5,-3){5}}
\put(34.5,0.5){\circle*{1}}
\put(39.5,0.5){\circle*{1}}
\put(33,5.7){\circle*{1}}
\put(41,5.7){\circle*{1}}
\put(37,8){\circle*{1}}
\put(37,-2){\makebox(0,0){$u_4(n)$}}
\end{picture}

\vspace*{5mm}
\caption{\footnotesize{Illustration of $f_4(n)$, $g_4(n)$, $h_4(n)$, $t_4(n)$, $s_4(n)$, $u_4(n)$. We only show the five outmost vertices explicitly, where each open circle is covered by a monomer while each solid circle is covered by a dimer.}} 
\label{fghtsu}
\end{figure}

$f_4(n)$, $g_4(n)$, $h_4(n)$, $t_4(n)$, $s_{4}(n)$ and $u_4(n)$ are illustrated in Fig. \ref{fghtsu}, where we only show the outmost vertices explicitly. Due to rotational symmetry, there are ${5\choose 1}=5$ orientations of $g_4(n)$, ${5\choose 2}=10$ orientations of $h_4(n)$, ${5\choose 3}=10$ orientations of $t_4(n)$, and ${5\choose 4}=5$ orientations of $s_{4}(n)$, such that
\beq
M_4(n) = f_4(n)+5g_4(n)+10h_4(n)+10t_4(n)+5s_4(n)+u_4(n) 
\label{meq4}
\eeq
for a non-negative integer $n$. The initial values at $n=0$ are $f_4(0)=1$, $g_4(0)=0$, $h_4(0)=1$, $t_4(0)=0$, $s_4(n)=3$, $u_4(n)=0$, and $M_4(0)=26$.

We write a program to get the recursion relations for $TH_{4}(n)$. These recursion relations are much more lengthier than those for $TH_{3}(n)$ and are omitted. The values of $M_4(n)$, $f_4(n)$, $g_4(n)$, $h_4(n)$, $t_4(n)$, $s_{4}(n)$ and $u_4(n)$ for $0 \le n \le 2$ are listed in Table \ref{table4d}.

\begin{table}[H]
\caption{\label{table4d} The values of $M_4(n)$, $f_4(n)$, $g_4(n)$, $h_4(n)$, $t_4(n)$, $s_{4}(n)$ and $u_4(n)$ with $0 \le n \le 2$.}
\begin{center}
\begin{tabular}{|c||r|r|r|}
\hline\hline 
$n$      & 0 &          1 &           2 \\ \hline\hline 
$M_4(n)$ & 26& 48,645,865 &  1,209,689,823,065,753,613,801,849,265,389,348,210,254 \\ \hline 
$f_4(n)$ & 1 &    510,980 &     12,567,379,442,065,248,794,102,222,711,306,394,841 \\ \hline 
$g_4(n)$ & 0 &    755,968 &     18,760,454,431,707,651,977,688,401,100,886,141,664 \\ \hline 
$h_4(n)$ & 1 &  1,123,642 &     28,005,432,734,266,093,414,497,192,140,551,929,071 \\ \hline 
$t_4(n)$ & 0 &  1,677,248 &     41,806,280,366,033,934,562,540,832,493,986,021,752 \\ \hline
$s_4(n)$ & 3 &  2,513,329 &     62,408,116,726,493,840,561,375,438,310,621,519,011 \\ \hline
$u_4(n)$ & 0 &  3,779,500 &     93,162,456,829,680,622,542,047,599,275,124,003,808 \\ \hline\hline 
\end{tabular}
\end{center}
\end{table}

For the four-dimensional Tower of Hanoi graph, let us define the ratios $\alpha_4(n)=\frac{f_4(n)}{g_4(n)}$ and $\omega_4(n)=\frac{s_4(n)}{u_4(n)}$, similar to those in Eq. (\ref{ratiodef}). It can be shown that $\alpha_4(n)$ decreases monotonically as $n$ increases while $\omega_4(n)$ increases monotonically, and 
\beq
\omega_4(n) < \frac{t_4(n)}{s_4(n)} < \frac{h_4(n)}{t_4(n)} < \frac{g_4(n)}{h_4(n)} < \alpha_4(n)
\eeq
for any positive integer $n$. The values of these ratios for $1 \le n \le 4$ are listed in Table \ref{tablesratio4}, and they approach to each other as $n$ increases. In the large $n$ limit, the numerical results give
\beqs
\lim_{n\rightarrow\infty} \alpha_4(n)&=&\lim_{n\rightarrow\infty}\frac{g_4(n)}{h_4(n)}=\lim_{n\rightarrow\infty}\frac{h_4(n)}{t_4(n)}=\lim_{n\rightarrow\infty}\frac{t_4(n)}{s_4(n)}=\lim_{n\rightarrow\infty}\omega_4(n)\cr
&=&0.66988575004174782028883689785... \
\eeqs

\begin{table}[H]
\caption{\label{tablesratio4} The values of $\alpha_4(n)$, $\omega_4(n)$, and other ratios with $1 \le n \le 4$. The last digits are rounded off.}
\begin{center}
\begin{tabular}{|c||r|r|r|r|r|}
\hline\hline 
$n$  &  1 &  2 & 3 & 4 \\ \hline\hline 
$\alpha_4(n)$   & 0.67592808161192 & 0.66988672837395 & 0.66988575004178 & 0.66988575004175 \\ \hline 
$g_4(n)/h_4(n)$ & 0.67278368021131 & 0.66988625420357 & 0.66988575004176 & 0.66988575004175 \\ \hline
$h_4(n)/t_4(n)$ & 0.66993193612394 & 0.66988578005661 & 0.66988575004175 & 0.66988575004175 \\ \hline 
$t_4(n)/s_4(n)$ & 0.66734120363868 & 0.66988530593307 & 0.66988575004173 & 0.66988575004175 \\ \hline
$\omega_4(n)$   & 0.66498981346739 & 0.66988483183294 & 0.66988575004172 & 0.66988575004175 \\ \hline\hline 
\end{tabular}
\end{center}
\end{table}

By the argument similar to that in Lemmas \ref{lemma3} and \ref{lemma4}, the entropy per site for dimer-monomers on $TH_4(n)$ is bounded:
\beq\label{z4d}
\frac{\ln{u_4(k)}}{5^{k+1}}+\frac{\ln{[1+2\omega_4(k)+2\omega_4(k)^2]}}{2(5^{k})} < z_{TH_4} < \frac{\ln{u_4(k)}}{5^{k+1}}+\frac{\ln{[1+2\alpha_4(k)+2\alpha_4(k)^2]}}{2(5^{k})} 
\eeq
where $k$ a positive integer. That leads to the following proposition.

\begin{propo}
The entropy per site for dimer-monomers on the four-dimensional Tower of Hanoi graph $TH_4(n)$ in the thermodynamic limit is
$z_{TH_4}=0.72291383087181938879...$
\end{propo}

The lower and upper bounds given in Eq. (\ref{z4d}) converge as rapid as that for the three-dimensional Tower of Hanoi graph $TH_3(n)$. The value of $z_{TH_4}$ is accurate to one hundred and twenty decimals when $k$ in Eq. (\ref{z4d}) is equal to six.

\section{Summary}

The lower and upper bounds of the entropy per site for dimer-monomers on $TH_2(n)$ in Ref. \cite{Dimer-monomer}, and on $TH_3(n)$, $TH_4(n)$ given above lead to the following conjecture for general $TH_d(n)$ with any dimension $d \ge 2$.

\begin{conj} 
Define $\alpha_d(n)$ to be the number of dimer-monomers on $TH_d(n)$ with all outmost vertices covered by monomers divided by the number with all but one outmost vertices covered by monomers. Define $\omega _d(n)$ to be the number of dimer-monomers on $TH_d(n)$ with all but one outmost vertices covered by dimers divided by the number with all outmost vertices covered by dimers. Define $\lambda _d(n)$ to be the number of dimer-monomers on $TH_d(n)$ with all outmost vertices covered by dimers. The entropy per site for dimer-monomers on the $d$-dimensional Tower of Hanoi graph $TH_d(n)$ is bounded:
\beqs
\frac{\ln{\lambda_d(k)}}{(d+1)^{k+1}}&+&\frac{\ln{[1+2\omega_d(k)+2\omega_d(k)^2]}}{2(d+1)^{k}}
< z_{TH_d} < \frac{\ln{\lambda_d(k)}}{(d+1)^{k+1}}+\frac{\ln{[1+2\alpha_d(k)+2\alpha_d(k)^2]}}{2(d+1)^{k}} \ . \cr &&
\eeqs
\end{conj}

Notice that although the lower and upper bounds given above are not exactly the same as those in \cite{Dimer-monomer}, the convergent rate is equivalent. The lower and upper bounds given above apply to $2 \le d \le 4$, and we conjecture that they are valid for any dimension $d$. It appears that the convergence of the lower and upper bounds of the entropy per site for dimer-monomers on $TH_d(n)$ becomes a bit faster as $d$ increases.

The present results can be compared with the entropy per site for dimer-monomers on the Sierpinski gasket (cf. \cite{sfs}), $z_{SG_d}$, as listed in Table \ref{tablesght}. For dimension $d=2, 3, 4$, the entropy per site on the Tower of Hanoi graph $TH_d$ is less than that on the Sierpinski gasket $SG_d$, and we conjecture that this relation remains true for arbitrary $d$. This can be attributed to the fact that the degree of $TH_d$ is less than that of $SG_d$.

\begin{table}[H]
\caption{\label{tablesght} The entropy per site for dimer-monomers on the Sierpinski gasket $SG_d$ and Tower of Hanoi graph $TH_d$ with $d=2, 3, 4$.}
\begin{center}
\begin{tabular}{|c|c|c|} \hline\hline 
$d$ & $z_{SG_d}$ & $z_{TH_d}$ \\ \hline\hline 
2   & 0.6562942369...  & 0.5764643016... \\ \hline
3   & 0.7811514674...  & 0.6571992114... \\ \hline
4   & 0.8767794029...  & 0.7229138308... \\ \hline\hline
\end{tabular}
\end{center}
\end{table}

\begin{acknowledgments}

This research of S.-C.C. was supported in part by the MOST grant 107-2515-S-006-002.

\end{acknowledgments}

\appendix

\section{Proof of the monotonicity of $\omega_3(n)$ and $\alpha_3(n)$}\label{appen1}

We shall show that $\omega_3(n)$ is an ascending function and $\alpha_3(n)$ is a descending function here.
Using $a,b,c$ to denote $\alpha_3(n)-\beta_3(n)$, $\beta_3(n)-\gamma_3(n)$, $\gamma_3(n)-\omega_3(n)$, respectively, and the definition given in Eq. (\ref{rationp1}), we find that $\omega_3(n+1)$ is always larger than $\omega_3(n)$ as follows.
\beqs
&&\omega_3(n+1)-\omega_3(n)
=\frac{\omega_3(n)}{E_3(n)}(D_3(n)-E_3(n))\cr
&=&\frac{\omega_3(n)}{E_3(n)}[(64a+64c+64b)\omega^{11}\cr
&&+(256ab+768bc+384b+512c^2+384c+256b^2+288a+512ac)\omega^{10}\cr
&&+(2688c^2+1792c^3+3584bc^2+1152c+2176b^2c+3840bc+2016ac+624a\cr
&&+384b^3+384ab^2+864ab+1792ac^2+1792abc+1104b+1152b^2)\omega^9\cr
&&+(2352b^2+7680b^2c^2+8064c^3+9120bc+5376abc^2+8960bc^3+256ab^3+256b^4\cr
&&+1968b+2304ab^2c+840a+5184abc+1152b^3+3584ac^3+6912c^2+1392ab\cr
&&+8064b^2c+864ab^2+2232c+6048ac^2+3584c^4+2560cb^3+3744ac+14976bc^2)\omega^8\cr
&&+(1392b^3+2808b^2+13296b^2c+13200bc+1416ab+288ab^3+9360ac^2+6960abc\cr
&&+4320ab^2c+912ab^2+1280ab^3c+384b^4+6144cb^3+64ab^4+1344b^4c+4224ac\cr
&&+64b^5+3072c+2400b+768a+11184c^2+17280c^3+13440c^4+29040bc^2+30720bc^3\cr
&&+23040b^2c^2+10080ac^3+12960abc^2+4480c^5+13440bc^4+14720b^2c^3+7040b^3c^2\cr
&&+4480ac^4+8960abc^3+5760ab^2c^2)\omega^7\cr
&&+(864b^3+2148b^2+12432b^2c+256ab^4c+12720bc+1020ab+144ab^3+8520ac^2\cr
&&+5736abc+3648ab^2c+600ab^2+1152ab^3c+144b^4+5760cb^3+1536b^4c+256b^5c\cr
&&+3168ac+3132c+2088b+492a+12432c^2+22440c^3+23040c^4+33216bc^2\cr
&&+47040bc^3+29664b^2c^2+12480ac^3+13920abc^2+3584c^6+13440c^5+12544bc^5\cr
&&+16640b^2c^4+10240b^3c^3+36480bc^4+34560b^2c^3+13056b^3c^2+2816b^4c^2+10080ac^4\cr
&&+17280abc^3+8640ab^2c^2+8960abc^4+7680ab^2c^3+2560ab^3c^2+3584ac^5)\omega^6\cr
&&+(384ab^4c^2+300b^3+1086b^2+7188b^2c+8436bc+552ab+24ab^3+4992ac^2\cr
&&+3252abc+1872ab^2c+252ab^2+432ab^3c+24b^4+2688cb^3+432b^4c+384b^5c^2\cr
&&+1638ac+2404c+1312b+220a+9792c^2+19008c^3+22560c^4+1792c^7\cr
&&+24192bc^2+7168bc^6+11136b^2c^5+8320b^3c^4+40224bc^3+20592b^2c^2+8640ac^3\cr
&&+8784abc^2+8064c^6+17280c^5+25344bc^5+28800b^2c^4+5376abc^5+5760ab^2c^4\cr
&&+2560ab^3c^3+13824b^3c^3+1792ac^6+2944b^4c^3+41520bc^4+32736b^2c^3+8928b^3c^2\cr
&&+2304b^4c^2+9360ac^4+13920abc^3+5472ab^2c^2+12960abc^4+8640ab^2c^3\cr
&&+1728ab^3c^2+6048ac^5)\omega^5\cr
&&+(256b^5c^3+1792abc^6+2304ab^2c^5+1280ab^3c^4+512ac^7+256ab^4c^3+60b^3\cr
&&+364b^2+2688b^2c+3870bc+226ab+1962ac^2+1326abc+600ab^2c+60ab^2\cr
&&+72ab^3c+744cb^3+72b^4c+590ac+1386c+588b+66a+5462c^2+10584c^3\cr
&&+13152c^4+2688c^7+11460bc^2+9600bc^6+12672b^2c^5+7296b^3c^4+20256bc^3\cr
&&+8364b^2c^2+512c^8+1536b^4c^4+2304bc^7+4096b^2c^6+3584b^3c^5+3648ac^3\cr
&&+3636abc^2+6912c^6+11400c^5+19104bc^5+17904b^2c^4+5184abc^5+4320ab^2c^4\cr
&&+1152ab^3c^3+6144b^3c^3+2016ac^6+1536b^4c^3+23856bc^4+15264b^2c^3+2880b^3c^2\cr
&&+432b^4c^2+4440ac^4+6096abc^3+2016ab^2c^2+6960abc^4+3648ab^2c^3+432ab^3c^2\cr
&&+3744ac^5)\omega^4\cr
&&+(864abc^6+864ab^2c^5+288ab^3c^4+288ac^7+6b^3+81b^2+670b^2c+1212bc\cr
&&+66ab+526ac^2+388abc+120ab^2c+6ab^2+120cb^3+147ac+588c+180b\cr
&&+12a+2094c^2+3718c^3+4320c^4+1152c^7+3518bc^2+3600bc^6+3888b^2c^5\cr
&&+1584b^3c^4+64ac^8+640b^2c^7+640b^3c^6+320b^4c^5+64b^5c^4+256abc^7+64c^9\cr
&&+320bc^8+5964bc^3+2118b^2c^2+384ab^2c^6+256ab^3c^5+64ab^4c^4+384c^8+384b^4c^4\cr
&&+1536bc^7+2304b^2c^6+1536b^3c^5+978ac^3+996abc^2+2352c^6+3648c^5+5712bc^5\cr
&&+4440b^2c^4+1392abc^5+912ab^2c^4+144ab^3c^3+1152b^3c^3+624ac^6+144b^4c^3\cr
&&+6816bc^4+3756b^2c^3+588b^3c^2+72b^4c^2+1152ac^4+1596abc^3+444ab^2c^2\cr
&&+1704abc^4+816ab^2c^3+72ab^3c^2+960ac^5)\omega^3\cr
&&+(12b^2+111b^2c+255bc+12ab+96ac^2+81abc+12ab^2c+12cb^3+24ac\cr
&&+175c+34b+a+513c^2+738c^3+666c^4+24c^7+684bc^2+96bc^6+144b^2c^5\cr
&&+96b^3c^4+978bc^3+324b^2c^2+162ac^3+174abc^2+144c^6+396c^5+432bc^5\cr
&&+432b^2c^4+72abc^5+72ab^2c^4+24ab^3c^3+144b^3c^3+24ac^6+24b^4c^3+852bc^4\cr
&&+516b^2c^3+60b^3c^2+162ac^4+222abc^3+60ab^2c^2+192abc^4+96ab^2c^3+96ac^5)\omega^2\cr
&&+(18b^2c^3+30b^2c^2+15ac^3+12ac^2+18bc^4+12abc+68c^2+33c+36bc+6c^5\cr
&&+b^2+30c^4+3b+75bc^2+12abc^3+2ac+ab+60bc^3+6ab^2c^2+15abc^2+6b^3c^2\cr
&&+63c^3+6ac^4+12b^2c)\omega\cr
&&+2bc^2+ac^2+b^2c+3c+3bc+abc+3c^2+c^3]>0 \ ,
\eeqs
where the inequality holds since all terms are positive. The relation $\alpha_3(n)-\alpha_3(n+1) > 0$ can be proved similarly, such that $\alpha_3(n)$ decreases monotonically as $n$ increases.

\end{document}